\documentclass[%
reprint,
groupedaddress,
amsmath,amssymb,
aps,
prx,
]{revtex4-2}

\usepackage{graphicx}
\usepackage{dcolumn}
\usepackage{bm}
\usepackage{hyperref}

\usepackage{ulem}
\usepackage{braket}
\usepackage{color}
\usepackage{xspace}
\usepackage{mhchem}

\newcommand{\hb}{\hat{b}}
\newcommand{\ldothatd}{\boldsymbol{\lambda}_\alpha\cdot\hat{\boldsymbol{D}}}
\newcommand{\ldotd}{\boldsymbol{\lambda}_\alpha\cdot\boldsymbol{D}}

\begin{document}

\preprint{QED-squeezed}

\title{
Light-Matter Hybridization and Entanglement from First-Principles}%

\author{Ilia M. Mazin}
\affiliation{
Theoretical Division, Los Alamos National Laboratory, Los Alamos, NM, 87544, USA.
}

\author{Yu Zhang}
\email{zhy@lanl.gov}
\affiliation{
Theoretical Division, Los Alamos National Laboratory, Los Alamos, NM, 87544, USA.
}

\date{\today}

\begin{abstract}
The hybridization between light and matter is fundamental for achieving cavity-induced control over quantum materials, necessitating accurate ab initio methods for their analysis. Among these, the quantum electrodynamics Hartree-Fock framework stands out as an essential mean field approximation for electron-electron and electron-photon interactions, forming the basis for advanced post-Hartree-Fock methods like quantum electrodynamics coupled cluster and auxiliary field quantum Monte Carlo. However, trivial quantum electrodynamics Hartere-Fock (QEDHF) methods assume a product state ansatz and thus cannot describe the light-matter Entanglement. Furthermore, our previous work on variational ansatz approaches lacked the capability to capture anharmonic or nonlinear fluctuations, limiting their applicability to strongly coupled systems. To overcome these limitations, we propose an extended QEDHF formalism by introducing a variational Squeeze transformation capable of describing anharmonic quantum fluctuations in photon fields. By optimizing the squeezing parameters, our framework provides a more nuanced and accurate characterization of photon field quantum fluctuations, significantly enhancing the predictive power of QEDHF in strong coupling regimes. Moreover, this formalism enhances the description of light-matter Entanglement, providing a first-principles framework for understanding light-matter hybridization and paving the way for deeper insights into cavity-modified quantum phenomena.
\end{abstract}

\maketitle

\section{Introduction}
\label{sec:intro}

Light-matter interactions are a fundamental area of research in quantum chemistry, condensed matter physics, and materials science~\cite{Forn-Diaz:2019vs}. Recent advancements in the strong coupling regime have led to the emergence of new frontiers in chemistry and materials science, such as polariton chemistry~\cite{Hutchison2012ACIE, Ebbesen2016ACR, Xiang:2024vc, Mandal:2023vh, Weight2023pccp} and cavity quantum materials~\cite{Schlawin:2022uq, Basov:2017ur, Ashida:2020vh, Hubener:2021wq}, which leverage strong light-matter interactions to modify materials and chemical properties. These include influencing chemical reactions~\cite{Zhang:2023vo, joel2022nc}, energy transfer~\cite{Timmer:2023wn}, condensation~\cite{Heinzen:2000uc}, quantum phase transitions~\cite{Budden:2021vz, Chiocchetta:2021ut}, and quantum information~\cite{Kavokin:2022tn, Ghosh:2020uk}. However, the mechanisms underlying light-induced states and emergent properties remain poorly understood.

\textit{Ab initio} methods are crucial for advancing our understanding of these complex systems, including both their thermodynamic and kinetic properties. These methods are also fundamental for nonadiabatic molecular dynamics simulations, which are vital for studying dynamical processes~\cite{Zhang2019JCP, Li2024:sn, Rana2023JPCA, luk_multiscale_2017, groenhof_tracking_2019, Li:2024tp}.
The challenges associated with accurately modeling light-matter interactions have stimulated the development of electronic structure theories for quantum electrodynamics (QED) problems, including QED density functional theory (QEDFT)~\cite{Ruggenthaler2014PRA, Tokatly2013PRL, Flick2017PNAS}, variational reduced density matrix methods~\cite{Matousek:2024vi}, and various wavefunction-based techniques. Among the wavefunction-based methods, the quantum electrodynamics Hartree-Fock (QEDHF) framework~\cite{haugland_coupled_2020, foley:2023CPR, Riso:2022uw, Zhang:2023vt, Cui:2024ul} stands out as a foundational mean-field theory, providing an essential first step for any first-principles approach. As a mean-field theory, QEDHF not only allows for a tractable approximation of the electron-photon interaction but also serves as the basis for more sophisticated post-Hartree-Fock (post-HF) methods, such as QED-coupled cluster (QED-CC)~\cite{haugland_coupled_2020, white:2020jcp, Weight2023pccp, PRR023262, yang:2024be}, Density Matrix Renormalization Group (DMRG)~\cite{Matousek:2024vi, Shaffer:2024et, Passetti:2023tc}, and quantum Monte Carlo (QMC)~\cite{Weight:2024vi, Weight:2024af}. Therefore, a robust QEDHF reference is critical, as it directly influences the accuracy and efficiency of these higher-order methods.

In our previous work, we developed a variational displacement transformation based QEDHF framework, namely VT-QEDHF~\cite{Zhang:2023vt}, enabling the treatment of light-matter interactions across all coupling regimes. The variational displaced QEDHF formalism effectively decouples the electron and photon degrees of freedom by shifting the photon field, providing a versatile tool for studying polaritonic states. However, the primary limitation of this approach lies in its focus on harmonic fluctuations, which, while adequate at lower coupling strengths, becomes insufficient as the coupling regime intensifies and anharmonic effects become significant.

To address this challenge, we introduce an extension to the QEDHF formalism by incorporating a variational squeeze (VSQ) transformation~\cite{Feinberg:1990sp, Trapper:1994tg, Barone:2008ug, Walls:1983tv}. The Squeeze transformation, widely utilized in quantum optics to describe states with modified uncertainty relations, captures the anharmonic fluctuations that arise in strongly coupled light-matter systems. Treating the squeezing parameters as variational allows us to optimize the system's energy, leading to a more accurate and comprehensive description of quantum fluctuations in the photon field. Furthermore, this scheme can quantitatively estimate light-matter Entanglement, as demonstrated by benchmarking against Density Matrix Renormalization Group (DMRG) calculations. Since Entanglement is an important indicator of polariton-induced properties, our approach provides significant insights into the interplay between electronic and photonic degrees of freedom. This advancement is expected to have substantial implications for the design and control of quantum materials and devices.

\section{Theory}
\label{sec:theory}

The total light-matter Hamiltonian of molecular quantum electrodynamics is commonly described using the nonrelativistic Pauli-Fierz (PF) Hamiltonian in the dipole approximation~\cite{CohenTannoudji1997, Mandal:2023vh, Weight2023pccp}:
\begin{align}\label{EQ:H_PF}
    \hat{H}_\mathrm{PF}
     =& \hat{H}_\mathrm{e} + \sum_{\alpha} \Big[ \omega_\alpha (\hat{b}^\dagger_\alpha\hat{b}_\alpha+\frac{1}{2})
    \nonumber\\
    &+\sqrt{\frac{\omega_\alpha}{2}} \boldsymbol{\lambda}_\alpha \cdot \hat{\boldsymbol{D}} (\hat{b}^\dagger_\alpha + \hat{b}_\alpha)
    + \frac{1}{2} (\boldsymbol{\lambda}_\alpha \cdot \hat{\boldsymbol{D}})^2 \Big].
\end{align}

Here, $\hat{H}_\mathrm{e} = \hat{T}_\mathrm{e} + \hat{V}$ represents the bare molecular Hamiltonian (excluding the nuclear kinetic operator) and includes all Coulomb interactions $\hat{V}$ between electrons and nuclei, as well as the electronic kinetic energy operators $\hat{T}_\mathrm{e}$. Explicitly, it can be written as
\begin{equation}
    \hat{H}_e = \sum_{\mu\nu}h_{\mu\nu} \hat{c}^\dagger_\mu \hat{c}_\nu +
    \frac{1}{2}\sum_{\mu\nu\lambda\sigma}I_{\mu\nu\lambda\sigma} \hat{c}^\dagger_\mu \hat{c}^\dagger_\lambda \hat{c}_\sigma \hat{c}_\nu.
\end{equation}
where $h$ and $I$ denote one-electron and two-electron integrals.
In Eq.~\ref{EQ:H_PF}, $\hat{\boldsymbol{D}}$ is the molecular dipole operator,
$\hat{\boldsymbol{D}}=\sum_i^{N_n} z_i\hat{\boldsymbol{R}}_i -\sum_i^{N_e} e\hat{\boldsymbol{r}}_i
\equiv\hat{\boldsymbol{D}}_n + \hat{\boldsymbol{D}}_e$,
which includes both electronic ($\hat{\boldsymbol{D}}_e$) and nuclear ($\hat{\boldsymbol{D}}_n$) components. The term $\boldsymbol{\lambda}_\alpha =\sqrt{\frac{1}{\epsilon_0 V}}\boldsymbol{e}_\alpha \equiv \lambda_\alpha \boldsymbol{e}_\alpha$ characterizes the coupling between the molecule and the cavity-quantized field. Here, $\omega_\alpha$ and $\boldsymbol{e}_\alpha$ represent the frequency and polarization of the electric field of cavity photon mode $\alpha$. The last term in Eq.~\ref{EQ:H_PF} describes the dipole self-energy (DSE), which ensures the Hamiltonian is bounded from below and exhibits the correct scaling with system size~\cite{Rokaj2018JPBAMOP}. Although the DSE term originates from the transverse component of the electric field and may vanish in certain nanoplasmonic cavities, the formalism presented here is general, irrespective of whether DSE is present or not. In this work, we will demonstrate applications of our methods to systems without DSE.

The eigenstate of the molecular QED Hamiltonian can be readily obtained by solving the time-independent Schr\"odinger equation
$\hat{H}_{\mathrm{PF}}\ket{\Psi} = E\ket{\Psi}$, 
where $\ket{\Psi}$ is the correlated electron-photon wavefunction, though the exact solution to the above quantum many-body equation is nontrivial.
Though there are many developments of ab initio theory for QED problems, most current work focuses on normal Fock or coherent state representation for the photonic DOFs~\cite{haugland_coupled_2020, foley:2023CPR}. Though the variational LF transformation-based QEDHF work significantly improves the mean-field solutions to the problems, it cannot capture the anharmonic photon quantum fluctuations. Here, we represent a new set of QEDHF theories in conjunction with squeezed photonic operators to address the problem. 

For convenience, we first define the following two operators,
\begin{align}
  \hat{D}(z) &=\prod_\alpha e^{-(z_\alpha \hat{b}_\alpha - h.c.)}, 
  \\
  \hat{S}(F) & = \prod_\alpha e^{\frac{1}{2}(F^*_\alpha\hat{b}^2_\alpha - F_\alpha\hat{b}^{\dag 2}_\alpha)}.
\end{align}
The displacement operator $\hat{D}(z)$ generates the coherent state $\ket{z}$ when acting on the vacuum state, where $z_\alpha \equiv \frac{ \boldsymbol{\lambda}_\alpha\cdot\langle\boldsymbol{D}\rangle}{\sqrt{2\omega_\alpha}}$ describes the displacement due to electron-photon coupling. Replacing $z$ with an operator $\hat{f}_\alpha \equiv \frac{f_\alpha \boldsymbol{\lambda}_\alpha \cdot \boldsymbol{D}}{\sqrt{2\omega_\alpha}}$ introduces a Gaussian ansatz with variational flexibility, capturing orbital relaxation effects as shown in our previous work~\cite{Zhang:2023vt}. The variational squeeze operator $\hat{S}(F_\alpha)$~\cite{Trapper:1994tg, Feinberg:1990sp}, where $F_\alpha = r_\alpha e^{i\theta}$, is widely used in quantum optics to describe states with modified uncertainty relations and has been instrumental in enhancing precision measurements, such as those in LIGO~\cite{Bailes:2021tz}. $F_\alpha$ is generally a complex number and is real in the atomic basis set without any gauge phase (i.e., $F=r$). 

As demonstrated in previous work, both coherent states ($\hat{D}(z)$) and the variational dispalcement ($\hat{D}(\hat{f})$) transformation have significantly improved the performance of QEDHF~\cite{Zhang:2023vt, foley:2023CPR}. Notably, the variational displacement transformation-based VT-QEDHF framework offers greater flexibility in optimizing orbital relaxation effects and provides a more robust reference across all coupling strengths. In this work, we further enhance QEDHF methods by introducing the variational squeeze (VSQ) operator, which encodes additional physical insights within an HF-like formalism. We will show that, with the inclusion of both the variational displacement transformation and squeeze operator, the VSQ-QEDHF solution transcends the mean-field approximation by producing wavefunctions that are not simple products of electronic and photonic degrees of freedom (DOFs). This advancement enables a more accurate description of Entanglement between electronic and photonic DOFs. 

Irrespective of the photonic DOF ans\"atze, the general QEDHF wavefunction (WF) ans\"atze for coupled molecular QED problems can be expressed as,
\begin{equation}
  \ket{\Psi}= \hat{U}\ket{\text{HF}}\otimes \ket{0_p},
\end{equation}
where $\ket{0_p}$ is the photonic vacuum state. $\ket{\text{HF}}=\exp\left[\sum_{pq}\xi_{pq}(\hat{b}^\dag_{p}\hat{b}_q - \hat{b}^\dag_q \hat{b}_p)\right]\prod^{N_o}_i \hat{b}^\dag_i\ket{0_e}$
is the conventional electronic HF ansatz. $\hat{U}$ defines the ansatz for the phtonic DOFs. 

Since the displacement operator (either $\hat{D}(z_\alpha)$ or $\hat{D}(\hat{f}_\alpha)$) and the VSQ operator $\hat{S}(F)$ do not commute, there are four possible combinations of the two operators, as shown in Table~\ref{table: ansatz}. However, as detailed in the SM, although $[\hat{S}(F), \hat{D}(z)] \neq 0$, the combinations $\hat{S}(F)\hat{D}(z)$ and $\hat{D}(z)\hat{S}(F)$ are equivalent up to a rotation in $F$. Consequently, only three unique ans\"atze arise from the four possible combinations.
Furthermore, as detailed in the SM, the Coherent Squeezed State (CSS) does not introduce any additional squeezing effect on the electronic Hamiltonian. While the squeeze operator renormalizes the bilinear term, its contribution to the total energy vanishes in the limit of a single-photon basis (vacuum state). Hence, it is straightforward to verify that the squeezed state does not affect the ground-state energy in the vacuum state.
In contrast, the Gaussian Squeezed State (GSS) ansatz significantly impacts the ground-state energy, even in the one-photon basis limit. Therefore, in the main text, we focus on using the variational GSS ansatz to further optimize the ground-state energy in the ``mean-field" limit. The SM explores the other three ans\"atze and their respective QEDHF formalisms.

\begin{table}[htb]
    \centering
    \caption{QEDHF ans\"atze with squeeze operators. Although $\hat{S}$ and $\hat{D}$ do not commute, some combinations are effectively equivalent (see SM for details).}
    \begin{tabular}{c|c}
    \hline\hline
      Squeezed coherent state (SCS)  & $\hat{U}(F,z) =\hat{S}(F)\hat{D}(z)$ \\ \hline
      Coherent Squeezed state (CSS)  & $\hat{U}(F,z) =\hat{D}(z)\hat{S}(F)$ \\ \hline
     Squeezed Gaussian state (SGS) & $\hat{U}(F,\hat{f}) =\hat{S}(F)\hat{D}(\hat{f})$  \\ \hline
     {\bf Gaussian Squeezed state (GSS)} & $\hat{U}(\hat{f},F) =\hat{D}(\hat{f})\hat{S}(F)$
     \\ \hline \hline
    \end{tabular}
    \label{table: ansatz}
\end{table}


Here, we derive the QEDHF formalism using the variational GSS ansatz for photons, which introduces two variational parameters to the $\hat{D}$ and $\hat{S}$ operators, respectively, to control the extent of the transformations,
\begin{equation}
    \ket{\Psi} = \hat{D}(\hat{f})\hat{S}(F)
    \ket{\text{HF}}\otimes \ket{0_p}
    \equiv \hat{U}(\hat{f},F)\ket{\text{HF}}\otimes \ket{0_p}.
\end{equation}
where $f_\alpha$ and $F_\alpha$ are the variational parameters in the coherent displacement and squeeze operators, respectively. Compared to the VT-QEDHF method~\cite{Zhang:2023vt}, the only difference is the introduction of the VSQ operator. The additional variational parameter $F < 0$ controls the squeezing of the wavefunctions~\cite{Feinberg:1990sp}.
The benefit of introducing a squeeze operator lies in its ability to partially capture the atomic fluctuations, which are absent in the VT-QEDHF method. The action of the unitary operator $\hat{\mathcal{U}} \equiv \hat{U}(\hat{f}, F)$ on the bosonic and electronic operators is detailed in the supplementary materials (SM).

The PF Hamiltonian after the $\hat{\mathcal{U}}\hat{H}_{PF}\hat{\mathcal{U}}$ transformation becomes,
\begin{align}\label{eq:hcs}
  \hat{\mathcal{H}}_{CS} = & \hat{\mathcal{H}}_e(\hat{\mathcal{X}})
  + \sum_{\alpha} e^{-r}\sqrt{\frac{\omega_\alpha}{2}}(\Delta\lambda_\alpha) \mathbf{e}_\alpha\cdot\mathbf{D}(\hb^\dag_\alpha+\hb_\alpha )
  \nonumber\\ &
  + \frac{(\Delta\lambda_\alpha)^2}{2}(\mathbf{e}_\alpha\cdot\mathbf{D})^2
  + \hat{\mathcal{H}}_{ph}.
\end{align}
This transformed Hamiltonian is formally similar to the original PF Hamiltonian but includes a dressed electronic Hamiltonian, renormalized residual bilinear coupling, residual DSE, and a dressed photonic Hamiltonian. A detailed derivation can be found in Sec.~\ref{sec:qedhfGSS} of the SM. In particular, the dressed photonic Hamiltonian $\hat{\mathcal{H}}_{ph}$ is
\begin{equation}
    \hat{\mathcal{H}}_{ph} = \omega_\alpha\left[\cosh(2r)
    (\hat{b}^\dag_\alpha \hat{b}_\alpha + \frac{1}{2})
    - \sinh(2r)(\hat{b}^2_\alpha + \hat{b}^{\dag2}_\alpha)\right].
\end{equation}

The dressed electronic Hamiltonian $\tilde{H}_e$ becomes, 
\begin{align}
  \hat{\mathcal{H}}_e(\hat{\mathcal{X}}) = \tilde{h}_{\mu\nu}\hat{E}_{\mu\nu} + \tilde{I}_{\mu\nu\lambda\sigma}\hat{E}_{\mu\nu\lambda\sigma}.
\end{align}
where 
\begin{align}
    \tilde{h}_{\mu\nu}&=\sum_{\mu'\nu} h_{\mu'\nu'}\mathcal{X}^\dag_{\mu\mu'}\mathcal{X}_{\nu\nu'}
    \\
    \tilde{I}_{\mu\nu\lambda\sigma}&= \sum_{\mu'\nu'\lambda'\sigma'}\mathcal{X}^\dag_{\mu\mu'}\mathcal{X}^\dag_{\nu\nu'}I_{\mu'\nu'\lambda'\sigma'} \mathcal{X}_{\lambda\lambda'}\mathcal{X}_{\sigma\sigma'}.
\end{align}
are the renormalized one- and two-body integrals. Here, $\hat{\mathcal{X}} = \exp\left[-f_\alpha e^{r}\frac{\ldothatd}{\sqrt{2\omega_\alpha}}(\hb^\dag_\alpha - \hb_\alpha)\right]$ is the displacement operator.
Compared to the variational displacement approach, the additional squeeze operator only renormalizes the bosonic Hamiltonian and dresses the one- and two-body integrals, leaving the bilinear coupling and DSE terms unchanged. As in the VT-QEDHF case, the $\mathcal{X}$ operators introduce Franck-Condon factors to the one-body and two-body integrals in the dipole basis, given for the one-body integrals as
$G^\alpha_{pq} = \exp\left[\frac{-f^2_\alpha (\eta_p - \eta_q)^2 e^{2r_\alpha}}{4\omega^2_\alpha}\right]$,
with a similar expression for the two-body integrals.

Applying the mean-field approximation to the transformed Hamiltonian yields the total energy functional for the VSQ-QEDHF method,\begin{align}\label{eq: totfunctional}
    E_{\text{tot}}
    = & \langle \hat{\mathcal{H}}_e(\mathcal{X})\rangle_{HF} +  \frac{(\Delta\lambda_\alpha)^2}{2} \bra{\mathrm{HF}} (\boldsymbol{e}_\alpha \cdot \hat{\boldsymbol{D}})^2 \ket{\mathrm{HF}}
    \nonumber\\ &
    + \omega_\alpha \cosh(2r)(n_\alpha+\frac{1}{2}).
\end{align}
Where the electronic energy functional $E_{e}\equiv \langle \hat{\mathcal{H}}_e(\mathcal{X})\rangle_{HF}$ becomes,
\begin{align}\label{eq: elec_functional}
    E_e = & \sum_{pq}\rho_{pq} h_{pq}G_{pq}
    \nonumber\\ & 
    + \sum_{pqrs}(2\rho_{pq}\rho_{rs}-\rho_{pr}\rho_{qs})I_{pqrs}G_{pqrs}.
\end{align}

Equations~\ref{eq: totfunctional} and~\ref{eq: elec_functional} represent the central results of this work. The electronic energy functional is formally identical to the standard Hartree-Fock (HF) method but incorporates renormalized one- and two-body integrals. The renormalization factors $G$ encode the effects of the displacement and squeeze operators.
Since $\cosh(2r) = \frac{e^{2r} + e^{-2r}}{2} \geq 1$, the lower bound of the photonic energy remains $E_\alpha \geq \omega_\alpha (n_\alpha + \frac{1}{2})$, ensuring that the Squeeze transformation does not alter the zero-point energy (ZPE) of the photonic subsystem. However, unlike the VT-QEDHF functional, the VSQ operator introduces additional renormalization factors in both the electronic energy functional (Eq.~\ref{eq: elec_functional}) and the photonic energy functional, accounting for nonlinear fluctuations in both subsystems.
This added flexibility allows for a more optimal balance between the contributions of the two terms, leading to a further reduction in the upper bound of the VSQ-QEDHF ground-state energy.

{\bf Fock matrix and gradient for SCF optimization.} As discussed earlier, the energy functional and corresponding Fock matrix for the SCS-based QEDHF are formally identical to those in the VT-QEDHF method~\cite{Zhang:2023vt}. However, an additional variational parameter, $r$, is introduced and must be minimized. The gradient of $E$ with respect to $r$ is given by
\begin{align}
    & \frac{\partial E}{\partial r} = \sinh(2r)\omega_\alpha
    - \sum_{pq} \rho_{pq}h_{pq}G_{pq}\frac{f^2_\alpha\xi_{pq}^2 e^{2r}}{2\omega^2_\alpha}
    \nonumber\\
    & - \sum_{pqrs}(2\rho_{pq}\rho_{rs} - \rho_{pr}\rho_{qs})I_{pqrs}G_{pqrs}\frac{f^2_\alpha\xi_{pqrs}^2 e^{2r}}{2\omega^2_\alpha}.
\end{align}
Here, $\xi_{pq}\equiv\eta_p - \eta_q$ and $\xi_{pqrs} = \eta_p + \eta_q - \eta_r - \eta_s$. Since the residual DSE remains unchanged under the squeezing transformation, it does not contribute to the gradient. Consequently, the SCF loop optimizes the squeezing parameter $r$ alongside the optimization of the density matrix and $f$.
The variational parameter $f$ generally represents the effective photon displacement due to electron-photon coupling, while the variational parameter $\tau \equiv e^{r}$ quantifies the anharmonic fluctuations of the photon wavefunctions.

{\bf Entropies and light-matter entanglements.} Conventional QEDHF, where the ansatz is a simple product of electronic and photonic components, does not capture light-matter Entanglement, which is a significant limitation in studying polaritonic systems. In contrast, both VT-QEDHF and VSQ-QEDHF introduce light-matter Entanglement, enabling a more accurate description of these interactions. This work explores the computation of light-matter Entanglement from first principles by employing the Von Neumann entropy as a measure.

The mean-field wavefunction of the transformed Hamiltonian is $\ket{\Phi} = \ket{HF}\otimes\ket{0_p}$. Hence, the ground state in the original AO-Fock basis is
\begin{align}
    \ket{\Psi}
     = & \hat{D}(\hat{f}) \ket{HF}\otimes \hat{S}(F)\ket{0} = \sum_\mu c_\mu \ket{\mu}\ket{z_\mu, F}.
\end{align}
More detailed derivation is given in Sec.~\ref{smsec:entropy} of the SM. The total density matrix can then be constructed from the wavefunction as $\rho = \ket{\Psi}\bra{\Psi}$. The density matrices (DMs) of the photonic subsystems are obtained by tracing over the electronic degrees of freedom (DOFs), respectively, i.e., $\rho_{ph} = \text{Tr}_e \ket{\Psi}\bra{\Psi}$. The corresponding entropy is given by $S = -k_B \text{Tr}[\rho_{ph}\ln \rho_{ph}]$, which is used to measure the strength of the light-matter Entanglement.

\section{Results and Discussion} 

To validate the VSQ-QEDHF method, implemented in the open-source package OpenMS~\cite{openms2023}, we applied it to both model and molecular systems to explore the physics of the squeezing effect. While some parameters may be challenging to achieve experimentally, a wide parameter space is used to highlight the differences between various QEDHF methods.

\subsection{One-dimensional Hubbard-Holstein model}
We begin with a one-dimensional Hubbard-Holstein-type model (at half-filling) with periodic boundary conditions to illustrate the physics of the squeeze operator in mean-field calculations. The model Hamiltonian is,
\begin{align}
  \hat{H} = & -t\sum_{ij} \hat{c}^\dag_i c_j  + \frac{U}{2}\sum_{ij} (\hat{n}_i - \frac{1}{2})(\hat{n}_j - \frac{1}{2})
  \nonumber\\
  & + \sqrt{\frac{\omega}{2}}g\sum_i \hat{n}_i (\hat{b}^\dag + \hat{b}) + \omega \hat{b}^\dag\hat{b}.
\end{align}
Here, $t$ is the nearest-neighbor hopping integral, and $U$ represents the local correlation strength. The coupling strength between the local dipole and the cavity photon is parameterized by $g$. The parameter $\Lambda = \frac{g^2}{D}$ characterizes the relative strength of the electron-photon coupling, where $D=2t$ is the half-bandwidth of the electronic density of states (DOS). The DSE term is omitted in this model, demonstrating that our implementation can handle systems with or without DSE.
Under the GSS transformation, the Hamiltonian becomes
\begin{align}
    \hat{H}_{G} = & -t\sum_{ij} \hat{\mathcal{X}}^\dag_i\hat{c}^\dag_i c_j  \hat{\mathcal{X}}_j
    + \frac{1}{2}\tilde{U}\sum_{ij} \hat{n}_i \hat{n}_j
    \nonumber\\
    & + \sqrt{\frac{\omega}{2}}\Delta g e^{-r} \hat{n}_i (\hat{b}^\dag + \hat{b})
    + \hat{\mathcal{H}}_{ph},
\end{align}
where $\tilde{U}=U-g^2f(2-f)$ is the effective electronic correlation after the transformation. This result indicates that the electron correlation is modified by light-matter interactions and can even become negative when $U \leq g^2 f(2-f)$, akin to the BCS electron pairing mechanism.

\begin{figure}
    \centering
    \includegraphics[width=0.97\linewidth]{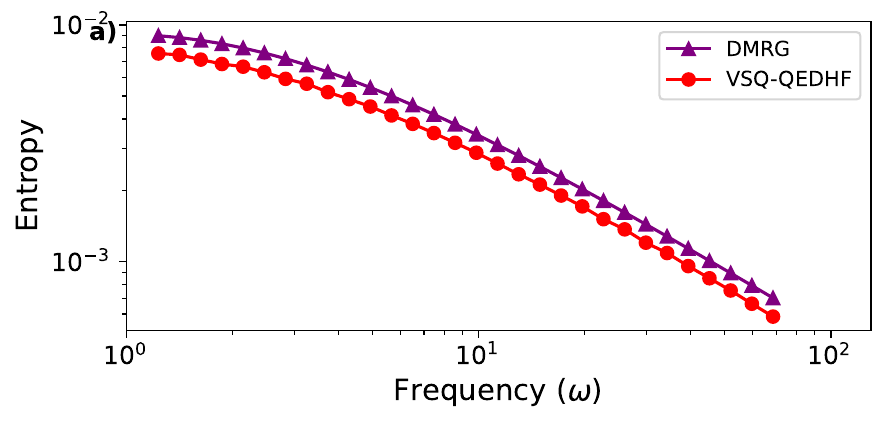} \\
    \vspace{-5pt}
    \includegraphics[width=1.0\linewidth]{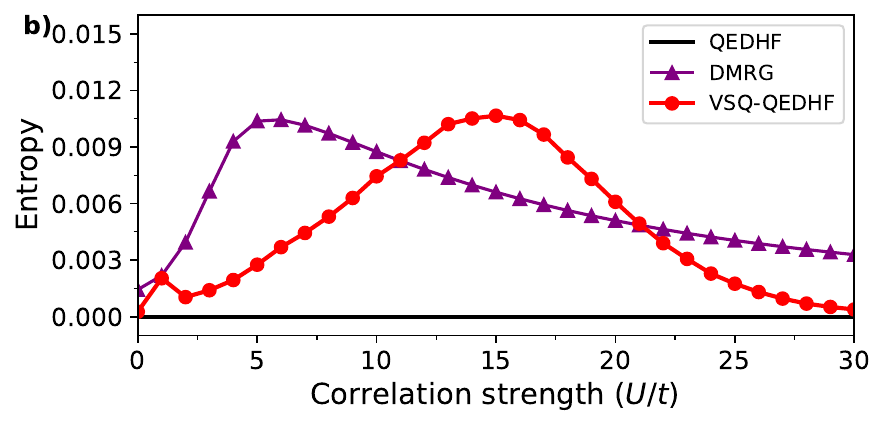}
    \vspace{-15pt}
    \caption{
    a) Comparison of variational on-site energies obtained from VT-QEDHF, VSQ-QEDHF, and DMRG.
    b) VSQ-QEDHF calculation of energy and light-matter Entanglement as a function of b) photon frequency $\omega$ and c) correlation strength $U$, with benchmarks against DMRG.
    Conventional QEDHF and other mean-field methods using decoupled ans\"atze yield qualitatively incorrect entropy dispersions in strongly correlated systems. In contrast, VSQ-QEDHF mitigates this issue with its advanced variational entangled ansatz.
    The kink observed in the VSQ-QEDHF curve at $U/t=1$ arises from discontinuities in the mean-field optimization.
    }
    \label{fig:model}
\end{figure}

We compute the light-matter Entanglement across different parameter spaces using the one-dimensional Hubbard-Holstein model. Calculations are performed on a 20-site system with periodic boundary conditions, assuming nearest-neighbor hopping and inter-site correlation. Coupling strength ($g$), photon frequency ($\omega$), and correlation strength ($U$) are expressed in units of the hopping parameter ($t$). For DMRG calculations, a bond dimension of 1,000 is used.

First, we examine the effect of photon frequency on light-matter Entanglement and benchmark the results against DMRG calculations performed using the TenPy package~\cite{tenpy}. The results are shown in Fig.~\ref{fig:model}a. While the VSQ-QEDHF method cannot describe electron-electron correlations, it accurately captures light-matter Entanglement. Benchmarking against DMRG calculations reveals quantitative agreement in computing light-matter Entanglement. This indicates that although the VSQ-QEDHF method underestimates electronic correlation, it effectively captures electron-photon correlation energies. This comparison underscores the importance of efficient photon ans\"atze for understanding electron-photon correlations.

Second, we analyze the behavior of Entanglement as a function of electron correlation strength, $U$. It is well known that in mean-field treatments decoupling light-matter interactions, the topology of light-matter Entanglement with respect to electron correlation is qualitatively incorrect~\cite{Passetti:2023tc}. By incorporating advanced electron-photon interactions through the VSQ formalism, the VSQ-QEDHF method achieves quantitative agreement with DMRG calculations in entropy dependence on electron correlation. As shown in Fig.~\ref{fig:model}b, light-matter Entanglement initially increases with increasing correlation strength, reaching a critical point before decreasing with a scaling of $1/U^2$. This reduction is attributed to strong correlation-induced localization, as illustrated in the inset of Fig.~\ref{fig:model}b. Mean-field approaches fail to capture this turnover, but the VSQ-QEDHF method reproduces it qualitatively. These benchmarks highlight the capability of the VSQ-QEDHF method to capture light-matter Entanglement across a wide range of correlation strengths.

\subsection{{First-principles calculations} }
\begin{figure}[!htb]
    \centering
    \includegraphics[width=0.99\linewidth]{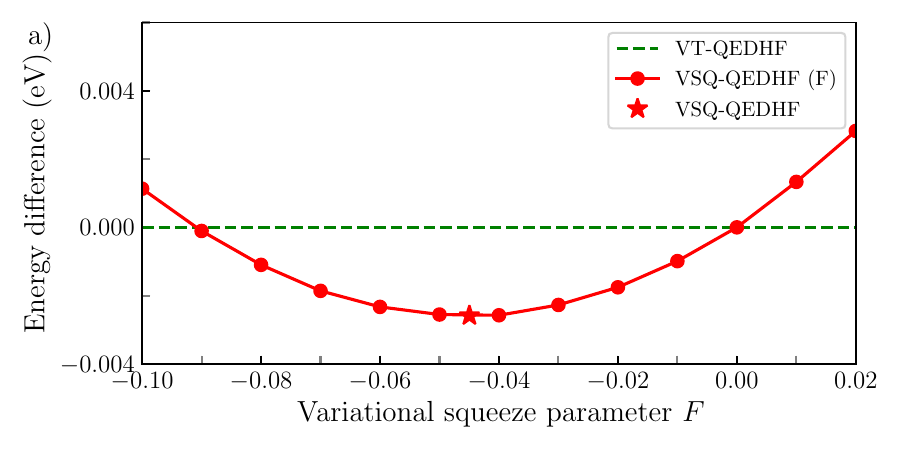} \\
    \vspace{-7pt}
    \includegraphics[width=0.99\linewidth]{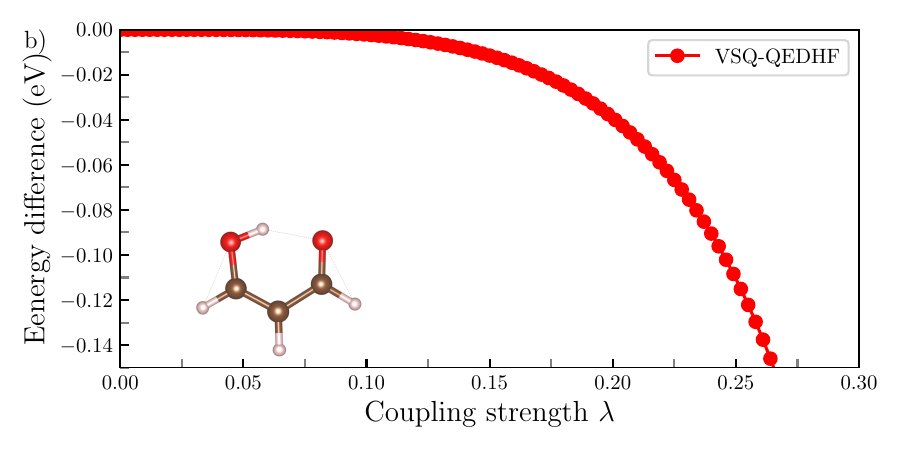} \\
    \vspace{-12pt}
    \caption{
    a) Energy difference ($E_{SQ} - E_{VT}$) as a function of the variational parameter $F$. When $F=0$, the VSQ-QEDHF method reduces to VT-QEDHF. The red star marks the VSQ-QEDHF result, where $F$ is optimized during the SCF procedure. The results highlight the power of the variational squeeze transformation, which further lowers the upper bound of the ground-state energy. {The parameters are $\lambda=0.1$, $\omega=0.05$}.
    b) Energy difference compared to VT-QEDHF as a function of coupling strength, illustrating the importance of the squeezing effect at large coupling limits. {$\omega=0.05$}
    }
    \label{fig:e_vs_f}
\end{figure}

Next, we demonstrate the implications of the variational squeeze operator in real molecular systems. As an example, we consider the \ce{C3H4O2} molecule, with its geometry depicted in Fig.~\ref{fig:e_vs_f}b). The 6-31G basis set is employed for this calculation. 

Fig.~\ref{fig:e_vs_f}a shows the energy difference between VSQ-QEDHF(F) (where $F$ is fixed) and VT-QEDHF as a function of the variational parameter $F$. The red star represents the VSQ-QEDHF energy, where $F$ is optimized during the SCF procedure alongside the density matrix. When $F=0$, the squeezing effect is absent, and the VSQ-QEDHF method reduces to the VT-QEDHF formalism, as confirmed by the numerical results. As $F$ decreases, the VSQ-QEDHF energy reaches a minimum, which is also achieved by the optimization procedure. This observation confirms the reliability of optimizing $F$ within the SCF loop.
This analysis highlights the relationship between VSQ-QEDHF and VT-QEDHF: VT-QEDHF corresponds to the non-squeezing limit of the VSQ-QEDHF method. The introduction of the squeeze operator captures additional nonlinear effects, further lowering the system's energy. This demonstrates the necessity of including the squeeze operator for accurately describing light-matter interactions, particularly at strong coupling.

As discussed above, the introduction of the squeeze operator incorporates a nonlinear, anharmonic effect on the photon field due to electron-photon interactions. This nonlinear behavior becomes increasingly significant as the coupling strength grows, particularly in the strong and ultrastrong coupling regimes. The squeeze operator modifies the photon states, leading to deformations in the vacuum state of the field that reflect a redistribution of quantum fluctuations. These fluctuations can no longer be described by a simple harmonic oscillator model, as they acquire a squeezed nature, altering both the amplitude and phase quadratures of the photon field.
Figure~\ref{fig:e_vs_f}b shows the energy difference between the VSQ-QEDHF and VT-QEDHF methods as a function of the coupling strength. The results indicate that the energy difference increases monotonically with increasing coupling strength, highlighting the growing importance of the squeeze operator in accurately capturing photon-mediated effects in this regime. At larger coupling strengths, the electron-photon interactions become so strong that the bare harmonic approximation (as employed in VT-QEDHF) fails to fully describe the system's quantum flutuation. The squeeze operator introduces critical corrections by accounting for photon number fluctuations and the resulting anharmonicity.
This correction is particularly significant because, in the ultrastrong coupling limit, these fluctuations influence both the ground and excited states of the system, leading to non-negligible energy shifts. By incorporating the squeeze operator, the VSQ-QEDHF method effectively captures these shifts, providing a more accurate description of the photon field and its coupling to the matter system.
Therefore, the squeezing effect is not a minor correction; it becomes an essential component in the ultrastrong coupling regime. Neglecting it would result in substantial inaccuracies in both energy estimations and the characterization of photon-matter Entanglement.

\begin{figure}[!htb]
  \centering
  \vspace{-7pt}
  \includegraphics[width=0.95\linewidth]{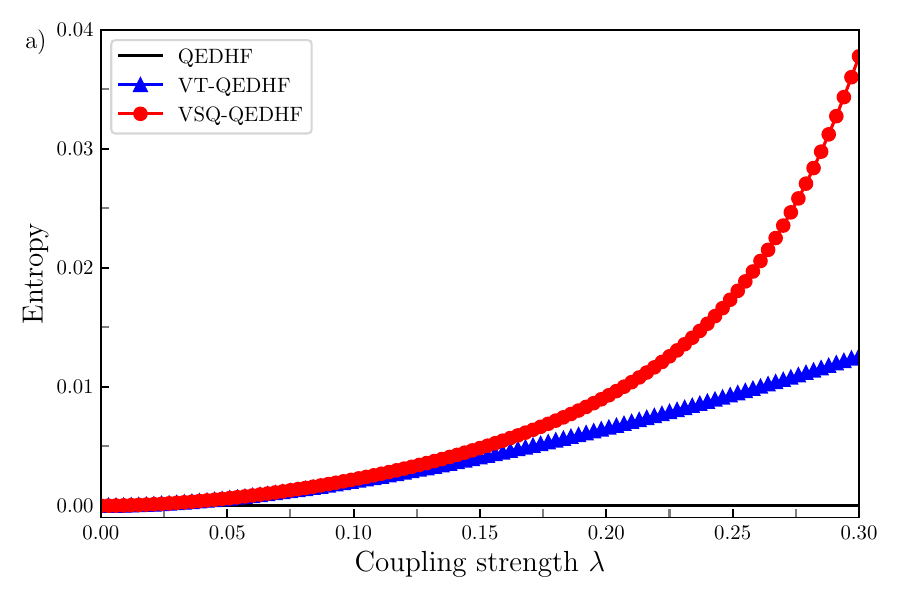}
  \\
  \vspace{-7pt}
  \includegraphics[width=0.95\linewidth]{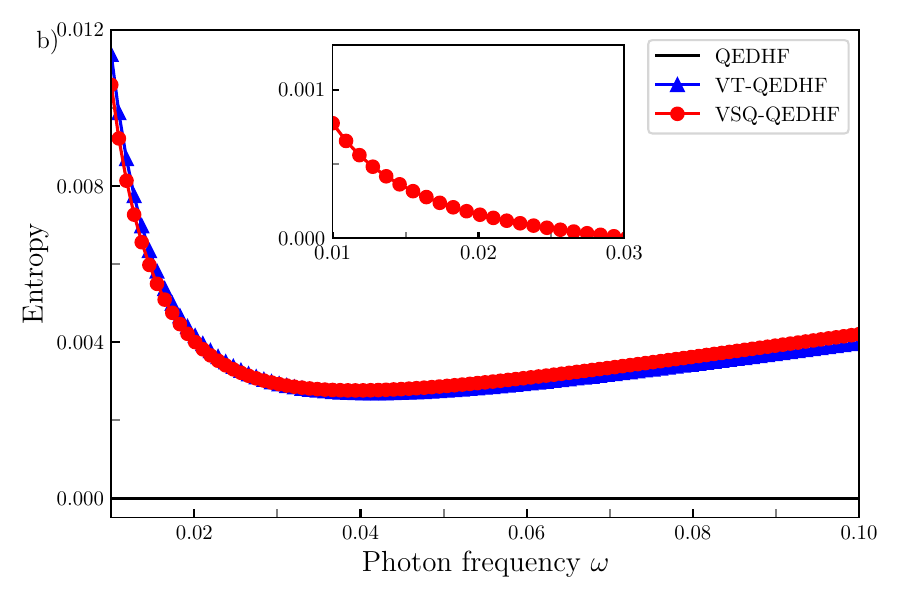}
  \\
  \vspace{-10pt}
  \caption{
  Entanglement as a function of coupling strength, $\lambda$, and photon frequency, $\omega$. The entropy difference increases with increasing coupling strength, emphasizing the importance of the squeezing operator in the strong coupling regime.
  In panels (a) and (b), the frequency and coupling strength are fixed at 0.05 and 0.1, respectively. The inset highlights the entropy difference between the VSQ-QEDHF and VT-QEDHF methods, which becomes more pronounced as the frequency decreases. Similarly, as shown in a), the difference grows with increasing coupling strength.
  }
  \label{fig:entropy_vs_w_lambda}
\end{figure}

Finally, we discuss the entropy and light-matter entanglement, which are essential for examining the quantum effects of cavity photons on electronic structures. As mentioned earlier, the trivial QEDHF ansatz assumes a simple product form, inherently failing to describe light-matter Entanglement. Consequently, the Von Neumann entropy is always zero, regardless of the frequency or coupling strength. In contrast, both VT-QEDHF and VSQ-QEDHF employ entangled ans\"atze, providing valuable insights into light-matter Entanglement across the parameter space. Consistent with the ground-state energy calculations, VSQ-QEDHF captures more electron-photon correlations, resulting in a more accurate estimation of light-matter Entanglement. 

Figure~\ref{fig:entropy_vs_w_lambda} illustrates the entropy as a function of cavity frequency $\omega$ and coupling strength $\lambda$. As shown in the figure, the entropy increases monotonically with increasing coupling strength, reflecting the strengthening of light-matter correlations. However, the relationship with frequency is more intricate. The entropy initially decreases with increasing frequency, reaches a minimum, then begins to rise, eventually saturating at higher frequency limits. This non-monotonic behavior highlights the delicate interplay between the photonic and electronic DOFs under varying frequency regimes. 

\section{\bf Summary}

In summary, this work introduces a significant extension of the QEDHF method through the variational squeezing ansatz. The QEDHF framework is a foundational mean-field theory that approximates electron-photon interactions. However, the conventional QEDHF method assumes a simple product state, limiting its ability to describe light-matter Entanglement. To address this limitation, we propose a Variational Squeeze (VSQ) transformation within the QEDHF framework. This extended formalism incorporates anharmonic fluctuations to more accurately capture the quantum behavior of the photon field, particularly in the strong coupling regime. By optimizing the squeezing parameters, the VSQ-QEDHF method provides a more comprehensive description of light-matter interactions, improving the predictive accuracy of QEDHF.

We demonstrated the efficacy of the VSQ-QEDHF method using a one-dimensional Hubbard-Holstein-type model and realistic molecular systems. The results show that the new approach better captures quantum fluctuations, leading to improved energy calculations. Additionally, the VSQ-QEDHF method effectively describes light-matter Entanglement, a capability lacking in traditional QEDHF methods. Our benchmark against DMRG indicates that the VSQ-QEDHF method exhibits qualitative agreement with DMRG results for entropy (light-matter Entanglement) as a function of coupling strength, photon frequency, and correlation strength. We observe that light-matter Entanglement decreases with increasing photon frequency. In terms of correlation strength, Entanglement initially increases, followed by a decrease as correlation-induced localization dominates at larger correlation limits. This nuanced behavior underscores the versatility of the VSQ-QEDHF method in capturing complex light-matter dynamics across varying interaction strengths.

The integration of the VSQ-QEDHF method with post-HF techniques holds great promise. This framework could accelerate the convergence of more advanced methods, such as QED-CC~\cite{haugland_coupled_2020, white:2020jcp, Weight2023pccp, PRR023262, yang:2024be} and the QED counterparts of Auxiliary Field or Diffusion Quantum Monte Carlo (AFQMC)~\cite{Weight:2024vi, Weber:2024af}, a project currently underway within our team. By incorporating variational squeezing, these post-HF methods are expected to benefit from more accurate initial conditions, enabling faster and more efficient simulations of complex light-matter systems. The ability to better account for anharmonic photon fluctuations and light-matter Entanglement should make these approaches more powerful and applicable across a wider range of coupling regimes, paving the way for precise explorations of quantum materials under strong light-matter coupling conditions.

\begin{acknowledgments}
We acknowledge support from the US DOE, Office of Science, Basic Energy Sciences, Chemical Sciences, Geosciences, and Biosciences Division under Triad National Security, LLC (``Triad'') contract Grant 89233218CNA000001 (FWP: LANLECF7). This research used computational resources provided by the Institutional Computing (IC) Program and the Darwin testbed at Los Alamos National Laboratory (LANL), funded by the Computational Systems and Software Environments subprogram of LANL's Advanced Simulation and Computing program. LANL is operated by Triad National Security, LLC, for the National Nuclear Security Administration of the US Department of Energy (Contract No. 89233218CNA000001).
\end{acknowledgments}


\bibliography{polaritonref}

\begin{thebibliography}{51}%
\makeatletter
\providecommand \@ifxundefined [1]{%
 \@ifx{#1\undefined}
}%
\providecommand \@ifnum [1]{%
 \ifnum #1\expandafter \@firstoftwo
 \else \expandafter \@secondoftwo
 \fi
}%
\providecommand \@ifx [1]{%
 \ifx #1\expandafter \@firstoftwo
 \else \expandafter \@secondoftwo
 \fi
}%
\providecommand \natexlab [1]{#1}%
\providecommand \enquote  [1]{``#1''}%
\providecommand \bibnamefont  [1]{#1}%
\providecommand \bibfnamefont [1]{#1}%
\providecommand \citenamefont [1]{#1}%
\providecommand \href@noop [0]{\@secondoftwo}%
\providecommand \href [0]{\begingroup \@sanitize@url \@href}%
\providecommand \@href[1]{\@@startlink{#1}\@@href}%
\providecommand \@@href[1]{\endgroup#1\@@endlink}%
\providecommand \@sanitize@url [0]{\catcode `\\12\catcode `\$12\catcode `\&12\catcode `\#12\catcode `\^12\catcode `\_12\catcode `\%12\relax}%
\providecommand \@@startlink[1]{}%
\providecommand \@@endlink[0]{}%
\providecommand \url  [0]{\begingroup\@sanitize@url \@url }%
\providecommand \@url [1]{\endgroup\@href {#1}{\urlprefix }}%
\providecommand \urlprefix  [0]{URL }%
\providecommand \Eprint [0]{\href }%
\providecommand \doibase [0]{https://doi.org/}%
\providecommand \selectlanguage [0]{\@gobble}%
\providecommand \bibinfo  [0]{\@secondoftwo}%
\providecommand \bibfield  [0]{\@secondoftwo}%
\providecommand \translation [1]{[#1]}%
\providecommand \BibitemOpen [0]{}%
\providecommand \bibitemStop [0]{}%
\providecommand \bibitemNoStop [0]{.\EOS\space}%
\providecommand \EOS [0]{\spacefactor3000\relax}%
\providecommand \BibitemShut  [1]{\csname bibitem#1\endcsname}%
\let\auto@bib@innerbib\@empty
\bibitem [{\citenamefont {Forn-D{\'\i}az}\ \emph {et~al.}(2019)\citenamefont {Forn-D{\'\i}az}, \citenamefont {Lamata}, \citenamefont {Rico}, \citenamefont {Kono},\ and\ \citenamefont {Solano}}]{Forn-Diaz:2019vs}%
  \BibitemOpen
  \bibfield  {author} {\bibinfo {author} {\bibfnamefont {P.}~\bibnamefont {Forn-D{\'\i}az}}, \bibinfo {author} {\bibfnamefont {L.}~\bibnamefont {Lamata}}, \bibinfo {author} {\bibfnamefont {E.}~\bibnamefont {Rico}}, \bibinfo {author} {\bibfnamefont {J.}~\bibnamefont {Kono}},\ and\ \bibinfo {author} {\bibfnamefont {E.}~\bibnamefont {Solano}},\ }\bibfield  {title} {\bibinfo {title} {Ultrastrong coupling regimes of light-matter interaction},\ }\href {https://doi.org/10.1103/RevModPhys.91.025005} {\bibfield  {journal} {\bibinfo  {journal} {Reviews of Modern Physics}\ }\textbf {\bibinfo {volume} {91}},\ \bibinfo {pages} {025005} (\bibinfo {year} {2019})}\BibitemShut {NoStop}%
\bibitem [{\citenamefont {Hutchison}\ \emph {et~al.}(2012)\citenamefont {Hutchison}, \citenamefont {Schwartz}, \citenamefont {Genet}, \citenamefont {Devaux},\ and\ \citenamefont {Ebbesen}}]{Hutchison2012ACIE}%
  \BibitemOpen
  \bibfield  {author} {\bibinfo {author} {\bibfnamefont {J.~A.}\ \bibnamefont {Hutchison}}, \bibinfo {author} {\bibfnamefont {T.}~\bibnamefont {Schwartz}}, \bibinfo {author} {\bibfnamefont {C.}~\bibnamefont {Genet}}, \bibinfo {author} {\bibfnamefont {E.}~\bibnamefont {Devaux}},\ and\ \bibinfo {author} {\bibfnamefont {T.~W.}\ \bibnamefont {Ebbesen}},\ }\bibfield  {title} {\bibinfo {title} {Modifying chemical landscapes by coupling to vacuum fields},\ }\href {https://doi.org/10.1002/anie.201107033} {\bibfield  {journal} {\bibinfo  {journal} {Angew. Chem. Int. Ed.}\ }\textbf {\bibinfo {volume} {51}},\ \bibinfo {pages} {1592} (\bibinfo {year} {2012})}\BibitemShut {NoStop}%
\bibitem [{\citenamefont {Ebbesen}(2016)}]{Ebbesen2016ACR}%
  \BibitemOpen
  \bibfield  {author} {\bibinfo {author} {\bibfnamefont {T.~W.}\ \bibnamefont {Ebbesen}},\ }\bibfield  {title} {\bibinfo {title} {Hybrid light{\textendash}matter states in a molecular and material science perspective},\ }\href {https://doi.org/10.1021/acs.accounts.6b00295} {\bibfield  {journal} {\bibinfo  {journal} {Acc. Chem. Res.}\ }\textbf {\bibinfo {volume} {49}},\ \bibinfo {pages} {2403} (\bibinfo {year} {2016})}\BibitemShut {NoStop}%
\bibitem [{\citenamefont {Xiang}\ and\ \citenamefont {Xiong}(2024)}]{Xiang:2024vc}%
  \BibitemOpen
  \bibfield  {author} {\bibinfo {author} {\bibfnamefont {B.}~\bibnamefont {Xiang}}\ and\ \bibinfo {author} {\bibfnamefont {W.}~\bibnamefont {Xiong}},\ }\bibfield  {title} {\bibinfo {title} {Molecular polaritons for chemistry, photonics and quantum technologies},\ }\href {https://doi.org/10.1021/acs.chemrev.3c00662} {\bibfield  {journal} {\bibinfo  {journal} {Chemical Reviews}\ }\textbf {\bibinfo {volume} {124}},\ \bibinfo {pages} {2512} (\bibinfo {year} {2024})}\BibitemShut {NoStop}%
\bibitem [{\citenamefont {Mandal}\ \emph {et~al.}(2023)\citenamefont {Mandal}, \citenamefont {Taylor}, \citenamefont {Weight}, \citenamefont {Koessler}, \citenamefont {Li},\ and\ \citenamefont {Huo}}]{Mandal:2023vh}%
  \BibitemOpen
  \bibfield  {author} {\bibinfo {author} {\bibfnamefont {A.}~\bibnamefont {Mandal}}, \bibinfo {author} {\bibfnamefont {M.~A.~D.}\ \bibnamefont {Taylor}}, \bibinfo {author} {\bibfnamefont {B.~M.}\ \bibnamefont {Weight}}, \bibinfo {author} {\bibfnamefont {E.~R.}\ \bibnamefont {Koessler}}, \bibinfo {author} {\bibfnamefont {X.}~\bibnamefont {Li}},\ and\ \bibinfo {author} {\bibfnamefont {P.}~\bibnamefont {Huo}},\ }\bibfield  {title} {\bibinfo {title} {Theoretical advances in polariton chemistry and molecular cavity quantum electrodynamics},\ }\href {https://doi.org/10.1021/acs.chemrev.2c00855} {\bibfield  {journal} {\bibinfo  {journal} {Chemical Reviews}\ }\textbf {\bibinfo {volume} {123}},\ \bibinfo {pages} {9786} (\bibinfo {year} {2023})}\BibitemShut {NoStop}%
\bibitem [{\citenamefont {Weight}\ \emph {et~al.}(2023)\citenamefont {Weight}, \citenamefont {Li},\ and\ \citenamefont {Zhang}}]{Weight2023pccp}%
  \BibitemOpen
  \bibfield  {author} {\bibinfo {author} {\bibfnamefont {B.}~\bibnamefont {Weight}}, \bibinfo {author} {\bibfnamefont {X.}~\bibnamefont {Li}},\ and\ \bibinfo {author} {\bibfnamefont {Y.}~\bibnamefont {Zhang}},\ }\bibfield  {title} {\bibinfo {title} {Theory and modeling of light-matter interactions in chemistry: current and future},\ }\href {https://doi.org/10.1039/D3CP01415K} {\bibfield  {journal} {\bibinfo  {journal} {Phys. Chem. Chem. Phys.}\ }\textbf {\bibinfo {volume} {25}},\ \bibinfo {pages} {31554} (\bibinfo {year} {2023})}\BibitemShut {NoStop}%
\bibitem [{\citenamefont {Schlawin}\ \emph {et~al.}(2022)\citenamefont {Schlawin}, \citenamefont {Kennes},\ and\ \citenamefont {Sentef}}]{Schlawin:2022uq}%
  \BibitemOpen
  \bibfield  {author} {\bibinfo {author} {\bibfnamefont {F.}~\bibnamefont {Schlawin}}, \bibinfo {author} {\bibfnamefont {D.~M.}\ \bibnamefont {Kennes}},\ and\ \bibinfo {author} {\bibfnamefont {M.~A.}\ \bibnamefont {Sentef}},\ }\bibfield  {title} {\bibinfo {title} {Cavity quantum materials},\ }\href {https://doi.org/10.1063/5.0083825} {\bibfield  {journal} {\bibinfo  {journal} {App. Phys. Rev.}\ }\textbf {\bibinfo {volume} {9}},\ \bibinfo {pages} {011312} (\bibinfo {year} {2022})}\BibitemShut {NoStop}%
\bibitem [{\citenamefont {Basov}\ \emph {et~al.}(2017)\citenamefont {Basov}, \citenamefont {Averitt},\ and\ \citenamefont {Hsieh}}]{Basov:2017ur}%
  \BibitemOpen
  \bibfield  {author} {\bibinfo {author} {\bibfnamefont {D.~N.}\ \bibnamefont {Basov}}, \bibinfo {author} {\bibfnamefont {R.~D.}\ \bibnamefont {Averitt}},\ and\ \bibinfo {author} {\bibfnamefont {D.}~\bibnamefont {Hsieh}},\ }\bibfield  {title} {\bibinfo {title} {Towards properties on demand in quantum materials},\ }\href {https://doi.org/10.1038/nmat5017} {\bibfield  {journal} {\bibinfo  {journal} {Nat. Mat.}\ }\textbf {\bibinfo {volume} {16}},\ \bibinfo {pages} {1077} (\bibinfo {year} {2017})}\BibitemShut {NoStop}%
\bibitem [{\citenamefont {Ashida}\ \emph {et~al.}(2020)\citenamefont {Ashida}, \citenamefont {{\.I}mamo{\u g}lu}, \citenamefont {Faist}, \citenamefont {Jaksch}, \citenamefont {Cavalleri},\ and\ \citenamefont {Demler}}]{Ashida:2020vh}%
  \BibitemOpen
  \bibfield  {author} {\bibinfo {author} {\bibfnamefont {Y.}~\bibnamefont {Ashida}}, \bibinfo {author} {\bibfnamefont {A.}~\bibnamefont {{\.I}mamo{\u g}lu}}, \bibinfo {author} {\bibfnamefont {J.}~\bibnamefont {Faist}}, \bibinfo {author} {\bibfnamefont {D.}~\bibnamefont {Jaksch}}, \bibinfo {author} {\bibfnamefont {A.}~\bibnamefont {Cavalleri}},\ and\ \bibinfo {author} {\bibfnamefont {E.}~\bibnamefont {Demler}},\ }\bibfield  {title} {\bibinfo {title} {Quantum electrodynamic control of matter: Cavity-enhanced ferroelectric phase transition},\ }\href {https://doi.org/10.1103/PhysRevX.10.041027} {\bibfield  {journal} {\bibinfo  {journal} {Phys. Rev. X}\ }\textbf {\bibinfo {volume} {10}},\ \bibinfo {pages} {041027} (\bibinfo {year} {2020})}\BibitemShut {NoStop}%
\bibitem [{\citenamefont {H{\"u}bener}\ \emph {et~al.}(2021)\citenamefont {H{\"u}bener}, \citenamefont {De~Giovannini}, \citenamefont {Sch{\"a}fer}, \citenamefont {Andberger}, \citenamefont {Ruggenthaler}, \citenamefont {Faist},\ and\ \citenamefont {Rubio}}]{Hubener:2021wq}%
  \BibitemOpen
  \bibfield  {author} {\bibinfo {author} {\bibfnamefont {H.}~\bibnamefont {H{\"u}bener}}, \bibinfo {author} {\bibfnamefont {U.}~\bibnamefont {De~Giovannini}}, \bibinfo {author} {\bibfnamefont {C.}~\bibnamefont {Sch{\"a}fer}}, \bibinfo {author} {\bibfnamefont {J.}~\bibnamefont {Andberger}}, \bibinfo {author} {\bibfnamefont {M.}~\bibnamefont {Ruggenthaler}}, \bibinfo {author} {\bibfnamefont {J.}~\bibnamefont {Faist}},\ and\ \bibinfo {author} {\bibfnamefont {A.}~\bibnamefont {Rubio}},\ }\bibfield  {title} {\bibinfo {title} {Engineering quantum materials with chiral optical cavities},\ }\href {https://doi.org/10.1038/s41563-020-00801-7} {\bibfield  {journal} {\bibinfo  {journal} {Nat. Mat.}\ }\textbf {\bibinfo {volume} {20}},\ \bibinfo {pages} {438} (\bibinfo {year} {2021})}\BibitemShut {NoStop}%
\bibitem [{\citenamefont {Zhang}\ \emph {et~al.}(2023)\citenamefont {Zhang}, \citenamefont {Nagata}, \citenamefont {Yao},\ and\ \citenamefont {Chin}}]{Zhang:2023vo}%
  \BibitemOpen
  \bibfield  {author} {\bibinfo {author} {\bibfnamefont {Z.}~\bibnamefont {Zhang}}, \bibinfo {author} {\bibfnamefont {S.}~\bibnamefont {Nagata}}, \bibinfo {author} {\bibfnamefont {K.-X.}\ \bibnamefont {Yao}},\ and\ \bibinfo {author} {\bibfnamefont {C.}~\bibnamefont {Chin}},\ }\bibfield  {title} {\bibinfo {title} {Many-body chemical reactions in a quantum degenerate gas},\ }\bibfield  {journal} {\bibinfo  {journal} {Nat. Phys.}\ }\href {https://doi.org/10.1038/s41567-023-02139-8} {10.1038/s41567-023-02139-8} (\bibinfo {year} {2023})\BibitemShut {NoStop}%
\bibitem [{\citenamefont {Pannir-Sivajothi}\ \emph {et~al.}(2022)\citenamefont {Pannir-Sivajothi}, \citenamefont {Campos-Gonzalez-Angulo}, \citenamefont {Martinez-Martinez}, \citenamefont {Sinha},\ and\ \citenamefont {Yuen-Zhou}}]{joel2022nc}%
  \BibitemOpen
  \bibfield  {author} {\bibinfo {author} {\bibfnamefont {S.}~\bibnamefont {Pannir-Sivajothi}}, \bibinfo {author} {\bibfnamefont {J.~A.}\ \bibnamefont {Campos-Gonzalez-Angulo}}, \bibinfo {author} {\bibfnamefont {L.~A.}\ \bibnamefont {Martinez-Martinez}}, \bibinfo {author} {\bibfnamefont {S.}~\bibnamefont {Sinha}},\ and\ \bibinfo {author} {\bibfnamefont {J.}~\bibnamefont {Yuen-Zhou}},\ }\bibfield  {title} {\bibinfo {title} {Driving chemical reactions with polariton condensates},\ }\href {https://doi.org/10.1038/s41467-022-29290-9} {\bibfield  {journal} {\bibinfo  {journal} {Nat. Commun.}\ }\textbf {\bibinfo {volume} {13}},\ \bibinfo {pages} {1645} (\bibinfo {year} {2022})}\BibitemShut {NoStop}%
\bibitem [{\citenamefont {Timmer}\ \emph {et~al.}(2023)\citenamefont {Timmer}, \citenamefont {Gittinger}, \citenamefont {Quenzel}, \citenamefont {Stephan}, \citenamefont {Zhang}, \citenamefont {Schumacher}, \citenamefont {L{\"u}tzen}, \citenamefont {Silies}, \citenamefont {Tretiak}, \citenamefont {Zhong}, \citenamefont {De~Sio},\ and\ \citenamefont {Lienau}}]{Timmer:2023wn}%
  \BibitemOpen
  \bibfield  {author} {\bibinfo {author} {\bibfnamefont {D.}~\bibnamefont {Timmer}}, \bibinfo {author} {\bibfnamefont {M.}~\bibnamefont {Gittinger}}, \bibinfo {author} {\bibfnamefont {T.}~\bibnamefont {Quenzel}}, \bibinfo {author} {\bibfnamefont {S.}~\bibnamefont {Stephan}}, \bibinfo {author} {\bibfnamefont {Y.}~\bibnamefont {Zhang}}, \bibinfo {author} {\bibfnamefont {M.~F.}\ \bibnamefont {Schumacher}}, \bibinfo {author} {\bibfnamefont {A.}~\bibnamefont {L{\"u}tzen}}, \bibinfo {author} {\bibfnamefont {M.}~\bibnamefont {Silies}}, \bibinfo {author} {\bibfnamefont {S.}~\bibnamefont {Tretiak}}, \bibinfo {author} {\bibfnamefont {J.-H.}\ \bibnamefont {Zhong}}, \bibinfo {author} {\bibfnamefont {A.}~\bibnamefont {De~Sio}},\ and\ \bibinfo {author} {\bibfnamefont {C.}~\bibnamefont {Lienau}},\ }\bibfield  {title} {\bibinfo {title} {Plasmon mediated coherent population oscillations in molecular aggregates},\ }\href {https://doi.org/10.1038/s41467-023-43578-4} {\bibfield  {journal} {\bibinfo  {journal} {Nat. Commun.}\
  }\textbf {\bibinfo {volume} {14}},\ \bibinfo {pages} {8035} (\bibinfo {year} {2023})}\BibitemShut {NoStop}%
\bibitem [{\citenamefont {Heinzen}\ \emph {et~al.}(2000)\citenamefont {Heinzen}, \citenamefont {Wynar}, \citenamefont {Drummond},\ and\ \citenamefont {Kheruntsyan}}]{Heinzen:2000uc}%
  \BibitemOpen
  \bibfield  {author} {\bibinfo {author} {\bibfnamefont {D.~J.}\ \bibnamefont {Heinzen}}, \bibinfo {author} {\bibfnamefont {R.}~\bibnamefont {Wynar}}, \bibinfo {author} {\bibfnamefont {P.~D.}\ \bibnamefont {Drummond}},\ and\ \bibinfo {author} {\bibfnamefont {K.~V.}\ \bibnamefont {Kheruntsyan}},\ }\bibfield  {title} {\bibinfo {title} {Superchemistry: Dynamics of coupled atomic and molecular bose-einstein condensates},\ }\href {https://doi.org/10.1103/PhysRevLett.84.5029} {\bibfield  {journal} {\bibinfo  {journal} {Phys. Rev. Lett.}\ }\textbf {\bibinfo {volume} {84}},\ \bibinfo {pages} {5029} (\bibinfo {year} {2000})}\BibitemShut {NoStop}%
\bibitem [{\citenamefont {Budden}\ \emph {et~al.}(2021)\citenamefont {Budden}, \citenamefont {Gebert}, \citenamefont {Buzzi}, \citenamefont {Jotzu}, \citenamefont {Wang}, \citenamefont {Matsuyama}, \citenamefont {Meier}, \citenamefont {Laplace}, \citenamefont {Pontiroli}, \citenamefont {Ricc{\`o}}, \citenamefont {Schlawin}, \citenamefont {Jaksch},\ and\ \citenamefont {Cavalleri}}]{Budden:2021vz}%
  \BibitemOpen
  \bibfield  {author} {\bibinfo {author} {\bibfnamefont {M.}~\bibnamefont {Budden}}, \bibinfo {author} {\bibfnamefont {T.}~\bibnamefont {Gebert}}, \bibinfo {author} {\bibfnamefont {M.}~\bibnamefont {Buzzi}}, \bibinfo {author} {\bibfnamefont {G.}~\bibnamefont {Jotzu}}, \bibinfo {author} {\bibfnamefont {E.}~\bibnamefont {Wang}}, \bibinfo {author} {\bibfnamefont {T.}~\bibnamefont {Matsuyama}}, \bibinfo {author} {\bibfnamefont {G.}~\bibnamefont {Meier}}, \bibinfo {author} {\bibfnamefont {Y.}~\bibnamefont {Laplace}}, \bibinfo {author} {\bibfnamefont {D.}~\bibnamefont {Pontiroli}}, \bibinfo {author} {\bibfnamefont {M.}~\bibnamefont {Ricc{\`o}}}, \bibinfo {author} {\bibfnamefont {F.}~\bibnamefont {Schlawin}}, \bibinfo {author} {\bibfnamefont {D.}~\bibnamefont {Jaksch}},\ and\ \bibinfo {author} {\bibfnamefont {A.}~\bibnamefont {Cavalleri}},\ }\bibfield  {title} {\bibinfo {title} {Evidence for metastable photo-induced superconductivity in k3c60},\ }\href {https://doi.org/10.1038/s41567-020-01148-1} {\bibfield
  {journal} {\bibinfo  {journal} {Nat. Phys.}\ }\textbf {\bibinfo {volume} {17}},\ \bibinfo {pages} {611} (\bibinfo {year} {2021})}\BibitemShut {NoStop}%
\bibitem [{\citenamefont {Chiocchetta}\ \emph {et~al.}(2021)\citenamefont {Chiocchetta}, \citenamefont {Kiese}, \citenamefont {Zelle}, \citenamefont {Piazza},\ and\ \citenamefont {Diehl}}]{Chiocchetta:2021ut}%
  \BibitemOpen
  \bibfield  {author} {\bibinfo {author} {\bibfnamefont {A.}~\bibnamefont {Chiocchetta}}, \bibinfo {author} {\bibfnamefont {D.}~\bibnamefont {Kiese}}, \bibinfo {author} {\bibfnamefont {C.~P.}\ \bibnamefont {Zelle}}, \bibinfo {author} {\bibfnamefont {F.}~\bibnamefont {Piazza}},\ and\ \bibinfo {author} {\bibfnamefont {S.}~\bibnamefont {Diehl}},\ }\bibfield  {title} {\bibinfo {title} {Cavity-induced quantum spin liquids},\ }\href {https://doi.org/10.1038/s41467-021-26076-3} {\bibfield  {journal} {\bibinfo  {journal} {Nat. Commun.}\ }\textbf {\bibinfo {volume} {12}},\ \bibinfo {pages} {5901} (\bibinfo {year} {2021})}\BibitemShut {NoStop}%
\bibitem [{\citenamefont {Kavokin}\ \emph {et~al.}(2022)\citenamefont {Kavokin}, \citenamefont {Liew}, \citenamefont {Schneider}, \citenamefont {Lagoudakis}, \citenamefont {Klembt},\ and\ \citenamefont {Hoefling}}]{Kavokin:2022tn}%
  \BibitemOpen
  \bibfield  {author} {\bibinfo {author} {\bibfnamefont {A.}~\bibnamefont {Kavokin}}, \bibinfo {author} {\bibfnamefont {T.~C.~H.}\ \bibnamefont {Liew}}, \bibinfo {author} {\bibfnamefont {C.}~\bibnamefont {Schneider}}, \bibinfo {author} {\bibfnamefont {P.~G.}\ \bibnamefont {Lagoudakis}}, \bibinfo {author} {\bibfnamefont {S.}~\bibnamefont {Klembt}},\ and\ \bibinfo {author} {\bibfnamefont {S.}~\bibnamefont {Hoefling}},\ }\bibfield  {title} {\bibinfo {title} {Polariton condensates for classical and quantum computing},\ }\href {https://doi.org/10.1038/s42254-022-00447-1} {\bibfield  {journal} {\bibinfo  {journal} {Nat. Rev. Phys.}\ }\textbf {\bibinfo {volume} {4}},\ \bibinfo {pages} {435} (\bibinfo {year} {2022})}\BibitemShut {NoStop}%
\bibitem [{\citenamefont {Ghosh}\ and\ \citenamefont {Liew}(2020)}]{Ghosh:2020uk}%
  \BibitemOpen
  \bibfield  {author} {\bibinfo {author} {\bibfnamefont {S.}~\bibnamefont {Ghosh}}\ and\ \bibinfo {author} {\bibfnamefont {T.~C.~H.}\ \bibnamefont {Liew}},\ }\bibfield  {title} {\bibinfo {title} {Quantum computing with exciton-polariton condensates},\ }\href {https://doi.org/10.1038/s41534-020-0244-x} {\bibfield  {journal} {\bibinfo  {journal} {npj Quantum Info.}\ }\textbf {\bibinfo {volume} {6}},\ \bibinfo {pages} {16} (\bibinfo {year} {2020})}\BibitemShut {NoStop}%
\bibitem [{\citenamefont {Zhang}\ \emph {et~al.}(2019)\citenamefont {Zhang}, \citenamefont {Nelson},\ and\ \citenamefont {Tretiak}}]{Zhang2019JCP}%
  \BibitemOpen
  \bibfield  {author} {\bibinfo {author} {\bibfnamefont {Y.}~\bibnamefont {Zhang}}, \bibinfo {author} {\bibfnamefont {T.}~\bibnamefont {Nelson}},\ and\ \bibinfo {author} {\bibfnamefont {S.}~\bibnamefont {Tretiak}},\ }\bibfield  {title} {\bibinfo {title} {Non-adiabatic molecular dynamics of molecules in the presence of strong light-matter interactions},\ }\href {https://doi.org/10.1063/1.5116550} {\bibfield  {journal} {\bibinfo  {journal} {J. Chem. Phys.}\ }\textbf {\bibinfo {volume} {151}},\ \bibinfo {pages} {154109} (\bibinfo {year} {2019})}\BibitemShut {NoStop}%
\bibitem [{\citenamefont {Li}\ \emph {et~al.}(2024{\natexlab{a}})\citenamefont {Li}, \citenamefont {Tretiak},\ and\ \citenamefont {Zhang}}]{Li2024:sn}%
  \BibitemOpen
  \bibfield  {author} {\bibinfo {author} {\bibfnamefont {X.}~\bibnamefont {Li}}, \bibinfo {author} {\bibfnamefont {S.}~\bibnamefont {Tretiak}},\ and\ \bibinfo {author} {\bibfnamefont {Y.}~\bibnamefont {Zhang}},\ }\bibfield  {title} {\bibinfo {title} {Semiclassical nonadiabatic molecular dynamics for molecular exciton-polaritons},\ }\href {https://doi.org/10.48550/arXiv.2410.16478} {\bibfield  {journal} {\bibinfo  {journal} {arXiv preprint arXiv:2410.16478}\ } (\bibinfo {year} {2024}{\natexlab{a}})}\BibitemShut {NoStop}%
\bibitem [{\citenamefont {Rana}\ \emph {et~al.}(2023)\citenamefont {Rana}, \citenamefont {Hohenstein},\ and\ \citenamefont {Mart{\'{i}}nez}}]{Rana2023JPCA}%
  \BibitemOpen
  \bibfield  {author} {\bibinfo {author} {\bibfnamefont {B.}~\bibnamefont {Rana}}, \bibinfo {author} {\bibfnamefont {E.~G.}\ \bibnamefont {Hohenstein}},\ and\ \bibinfo {author} {\bibfnamefont {T.~J.}\ \bibnamefont {Mart{\'{i}}nez}},\ }\bibfield  {title} {\bibinfo {title} {Simulating the excited-state dynamics of polaritons with ab initio multiple spawning},\ }\href {https://doi.org/10.1021/acs.jpca.3c06607} {\bibfield  {journal} {\bibinfo  {journal} {J. Phys. Chem. A}\ }\textbf {\bibinfo {volume} {128}},\ \bibinfo {pages} {139} (\bibinfo {year} {2023})}\BibitemShut {NoStop}%
\bibitem [{\citenamefont {Luk}\ \emph {et~al.}(2017)\citenamefont {Luk}, \citenamefont {Feist}, \citenamefont {Toppari},\ and\ \citenamefont {Groenhof}}]{luk_multiscale_2017}%
  \BibitemOpen
  \bibfield  {author} {\bibinfo {author} {\bibfnamefont {H.~L.}\ \bibnamefont {Luk}}, \bibinfo {author} {\bibfnamefont {J.}~\bibnamefont {Feist}}, \bibinfo {author} {\bibfnamefont {J.~J.}\ \bibnamefont {Toppari}},\ and\ \bibinfo {author} {\bibfnamefont {G.}~\bibnamefont {Groenhof}},\ }\bibfield  {title} {\bibinfo {title} {Multiscale {Molecular} {Dynamics} {Simulations} of {Polaritonic} {Chemistry}},\ }\href {https://doi.org/10.1021/acs.jctc.7b00388} {\bibfield  {journal} {\bibinfo  {journal} {J. Chem. Theory Comput.}\ }\textbf {\bibinfo {volume} {13}},\ \bibinfo {pages} {4324} (\bibinfo {year} {2017})}\BibitemShut {NoStop}%
\bibitem [{\citenamefont {Groenhof}\ \emph {et~al.}(2019)\citenamefont {Groenhof}, \citenamefont {Climent}, \citenamefont {Feist}, \citenamefont {Morozov},\ and\ \citenamefont {Toppari}}]{groenhof_tracking_2019}%
  \BibitemOpen
  \bibfield  {author} {\bibinfo {author} {\bibfnamefont {G.}~\bibnamefont {Groenhof}}, \bibinfo {author} {\bibfnamefont {C.}~\bibnamefont {Climent}}, \bibinfo {author} {\bibfnamefont {J.}~\bibnamefont {Feist}}, \bibinfo {author} {\bibfnamefont {D.}~\bibnamefont {Morozov}},\ and\ \bibinfo {author} {\bibfnamefont {J.~J.}\ \bibnamefont {Toppari}},\ }\bibfield  {title} {\bibinfo {title} {Tracking {Polariton} {Relaxation} with {Multiscale} {Molecular} {Dynamics} {Simulations}},\ }\href {https://doi.org/10.1021/acs.jpclett.9b02192} {\bibfield  {journal} {\bibinfo  {journal} {J. Phys. Chem. Lett.}\ }\textbf {\bibinfo {volume} {10}},\ \bibinfo {pages} {5476} (\bibinfo {year} {2019})}\BibitemShut {NoStop}%
\bibitem [{\citenamefont {Li}\ \emph {et~al.}(2024{\natexlab{b}})\citenamefont {Li}, \citenamefont {Lubbers}, \citenamefont {Tretiak}, \citenamefont {Barros},\ and\ \citenamefont {Zhang}}]{Li:2024tp}%
  \BibitemOpen
  \bibfield  {author} {\bibinfo {author} {\bibfnamefont {X.}~\bibnamefont {Li}}, \bibinfo {author} {\bibfnamefont {N.}~\bibnamefont {Lubbers}}, \bibinfo {author} {\bibfnamefont {S.}~\bibnamefont {Tretiak}}, \bibinfo {author} {\bibfnamefont {K.}~\bibnamefont {Barros}},\ and\ \bibinfo {author} {\bibfnamefont {Y.}~\bibnamefont {Zhang}},\ }\bibfield  {title} {\bibinfo {title} {Machine learning framework for modeling exciton polaritons in molecular materials},\ }\href {https://doi.org/10.1021/acs.jctc.3c01068} {\bibfield  {journal} {\bibinfo  {journal} {J. Chem. Theory Comput.}\ }\textbf {\bibinfo {volume} {20}},\ \bibinfo {pages} {891} (\bibinfo {year} {2024}{\natexlab{b}})}\BibitemShut {NoStop}%
\bibitem [{\citenamefont {Ruggenthaler}\ \emph {et~al.}(2014)\citenamefont {Ruggenthaler}, \citenamefont {Flick}, \citenamefont {Pellegrini}, \citenamefont {Appel}, \citenamefont {Tokatly},\ and\ \citenamefont {Rubio}}]{Ruggenthaler2014PRA}%
  \BibitemOpen
  \bibfield  {author} {\bibinfo {author} {\bibfnamefont {M.}~\bibnamefont {Ruggenthaler}}, \bibinfo {author} {\bibfnamefont {J.}~\bibnamefont {Flick}}, \bibinfo {author} {\bibfnamefont {C.}~\bibnamefont {Pellegrini}}, \bibinfo {author} {\bibfnamefont {H.}~\bibnamefont {Appel}}, \bibinfo {author} {\bibfnamefont {I.~V.}\ \bibnamefont {Tokatly}},\ and\ \bibinfo {author} {\bibfnamefont {A.}~\bibnamefont {Rubio}},\ }\bibfield  {title} {\bibinfo {title} {Quantum-electrodynamical density-functional theory: Bridging quantum optics and electronic-structure theory},\ }\href {https://doi.org/10.1103/physreva.90.012508} {\bibfield  {journal} {\bibinfo  {journal} {Phys. Rev. A}\ }\textbf {\bibinfo {volume} {90}},\ \bibinfo {pages} {012508} (\bibinfo {year} {2014})}\BibitemShut {NoStop}%
\bibitem [{\citenamefont {Tokatly}(2013)}]{Tokatly2013PRL}%
  \BibitemOpen
  \bibfield  {author} {\bibinfo {author} {\bibfnamefont {I.~V.}\ \bibnamefont {Tokatly}},\ }\bibfield  {title} {\bibinfo {title} {Time-dependent density functional theory for many-electron systems interacting with cavity photons},\ }\href {https://doi.org/10.1103/physrevlett.110.233001} {\bibfield  {journal} {\bibinfo  {journal} {Phys. Rev. Lett.}\ }\textbf {\bibinfo {volume} {110}},\ \bibinfo {pages} {233001} (\bibinfo {year} {2013})}\BibitemShut {NoStop}%
\bibitem [{\citenamefont {Flick}\ \emph {et~al.}(2017)\citenamefont {Flick}, \citenamefont {Ruggenthaler}, \citenamefont {Appel},\ and\ \citenamefont {Rubio}}]{Flick2017PNAS}%
  \BibitemOpen
  \bibfield  {author} {\bibinfo {author} {\bibfnamefont {J.}~\bibnamefont {Flick}}, \bibinfo {author} {\bibfnamefont {M.}~\bibnamefont {Ruggenthaler}}, \bibinfo {author} {\bibfnamefont {H.}~\bibnamefont {Appel}},\ and\ \bibinfo {author} {\bibfnamefont {A.}~\bibnamefont {Rubio}},\ }\bibfield  {title} {\bibinfo {title} {Atoms and molecules in cavities, from weak to strong coupling in quantum-electrodynamics ({QED}) chemistry},\ }\href {https://doi.org/10.1073/pnas.1615509114} {\bibfield  {journal} {\bibinfo  {journal} {Proc. Natl. Acad. Sci.}\ }\textbf {\bibinfo {volume} {114}},\ \bibinfo {pages} {3026} (\bibinfo {year} {2017})}\BibitemShut {NoStop}%
\bibitem [{\citenamefont {Matou{\v s}ek}\ \emph {et~al.}(2024)\citenamefont {Matou{\v s}ek}, \citenamefont {Vu}, \citenamefont {Govind}, \citenamefont {Foley},\ and\ \citenamefont {Veis}}]{Matousek:2024vi}%
  \BibitemOpen
  \bibfield  {author} {\bibinfo {author} {\bibfnamefont {M.}~\bibnamefont {Matou{\v s}ek}}, \bibinfo {author} {\bibfnamefont {N.}~\bibnamefont {Vu}}, \bibinfo {author} {\bibfnamefont {N.}~\bibnamefont {Govind}}, \bibinfo {author} {\bibfnamefont {J.~J.~I.}\ \bibnamefont {Foley}},\ and\ \bibinfo {author} {\bibfnamefont {L.}~\bibnamefont {Veis}},\ }\bibfield  {title} {\bibinfo {title} {Polaritonic chemistry using the density matrix renormalization group method},\ }\href {https://doi.org/10.1021/acs.jctc.4c00986} {\bibfield  {journal} {\bibinfo  {journal} {J. Chem. Theory Comput.}\ }\textbf {\bibinfo {volume} {20}},\ \bibinfo {pages} {9424} (\bibinfo {year} {2024})}\BibitemShut {NoStop}%
\bibitem [{\citenamefont {Haugland}\ \emph {et~al.}(2020)\citenamefont {Haugland}, \citenamefont {Ronca}, \citenamefont {Kj{\o}nstad}, \citenamefont {Rubio},\ and\ \citenamefont {Koch}}]{haugland_coupled_2020}%
  \BibitemOpen
  \bibfield  {author} {\bibinfo {author} {\bibfnamefont {T.~S.}\ \bibnamefont {Haugland}}, \bibinfo {author} {\bibfnamefont {E.}~\bibnamefont {Ronca}}, \bibinfo {author} {\bibfnamefont {E.~F.}\ \bibnamefont {Kj{\o}nstad}}, \bibinfo {author} {\bibfnamefont {A.}~\bibnamefont {Rubio}},\ and\ \bibinfo {author} {\bibfnamefont {H.}~\bibnamefont {Koch}},\ }\bibfield  {title} {\bibinfo {title} {Coupled cluster theory for molecular polaritons: Changing ground and excited states},\ }\href {https://doi.org/10.1103/PhysRevX.10.041043} {\bibfield  {journal} {\bibinfo  {journal} {Phys. Rev. X}\ }\textbf {\bibinfo {volume} {10}},\ \bibinfo {pages} {041043} (\bibinfo {year} {2020})}\BibitemShut {NoStop}%
\bibitem [{\citenamefont {Foley}\ \emph {et~al.}(2023)\citenamefont {Foley}, \citenamefont {McTague},\ and\ \citenamefont {DePrince}}]{foley:2023CPR}%
  \BibitemOpen
  \bibfield  {author} {\bibinfo {author} {\bibfnamefont {I.}~\bibnamefont {Foley}, \bibfnamefont {Jonathan~J.}}, \bibinfo {author} {\bibfnamefont {J.~F.}\ \bibnamefont {McTague}},\ and\ \bibinfo {author} {\bibfnamefont {I.}~\bibnamefont {DePrince}, \bibfnamefont {A.~Eugene}},\ }\bibfield  {title} {\bibinfo {title} {{Ab initio methods for polariton chemistry}},\ }\href {https://doi.org/10.1063/5.0167243} {\bibfield  {journal} {\bibinfo  {journal} {Chem. Phys. Rev.}\ }\textbf {\bibinfo {volume} {4}},\ \bibinfo {pages} {041301} (\bibinfo {year} {2023})}\BibitemShut {NoStop}%
\bibitem [{\citenamefont {Riso}\ \emph {et~al.}(2022)\citenamefont {Riso}, \citenamefont {Haugland}, \citenamefont {Ronca},\ and\ \citenamefont {Koch}}]{Riso:2022uw}%
  \BibitemOpen
  \bibfield  {author} {\bibinfo {author} {\bibfnamefont {R.~R.}\ \bibnamefont {Riso}}, \bibinfo {author} {\bibfnamefont {T.~S.}\ \bibnamefont {Haugland}}, \bibinfo {author} {\bibfnamefont {E.}~\bibnamefont {Ronca}},\ and\ \bibinfo {author} {\bibfnamefont {H.}~\bibnamefont {Koch}},\ }\bibfield  {title} {\bibinfo {title} {Molecular orbital theory in cavity qed environments},\ }\href {https://doi.org/10.1038/s41467-022-29003-2} {\bibfield  {journal} {\bibinfo  {journal} {Nat. Commun.}\ }\textbf {\bibinfo {volume} {13}},\ \bibinfo {pages} {1368} (\bibinfo {year} {2022})}\BibitemShut {NoStop}%
\bibitem [{\citenamefont {Li}\ and\ \citenamefont {Zhang}(2023)}]{Zhang:2023vt}%
  \BibitemOpen
  \bibfield  {author} {\bibinfo {author} {\bibfnamefont {X.}~\bibnamefont {Li}}\ and\ \bibinfo {author} {\bibfnamefont {Y.}~\bibnamefont {Zhang}},\ }\bibfield  {title} {\bibinfo {title} {First-principles molecular quantum electrodynamics theory at all coupling strengths},\ }\href {https://doi.org/10.48550/arXiv.2310.18228} {\bibfield  {journal} {\bibinfo  {journal} {arXiv:2310.18228}\ } (\bibinfo {year} {2023})}\BibitemShut {NoStop}%
\bibitem [{\citenamefont {Cui}\ \emph {et~al.}(2024)\citenamefont {Cui}, \citenamefont {Mandal},\ and\ \citenamefont {Reichman}}]{Cui:2024ul}%
  \BibitemOpen
  \bibfield  {author} {\bibinfo {author} {\bibfnamefont {Z.-H.}\ \bibnamefont {Cui}}, \bibinfo {author} {\bibfnamefont {A.}~\bibnamefont {Mandal}},\ and\ \bibinfo {author} {\bibfnamefont {D.~R.}\ \bibnamefont {Reichman}},\ }\bibfield  {title} {\bibinfo {title} {Variational lang--firsov approach plus m{\o}ller--plesset perturbation theory with applications to ab initio polariton chemistry},\ }\href {https://doi.org/10.1021/acs.jctc.3c01166} {\bibfield  {journal} {\bibinfo  {journal} {J. Chem. Theory Comput.}\ }\textbf {\bibinfo {volume} {20}},\ \bibinfo {pages} {1143} (\bibinfo {year} {2024})}\BibitemShut {NoStop}%
\bibitem [{\citenamefont {White}\ \emph {et~al.}(2020)\citenamefont {White}, \citenamefont {Gao}, \citenamefont {Minnich},\ and\ \citenamefont {Chan}}]{white:2020jcp}%
  \BibitemOpen
  \bibfield  {author} {\bibinfo {author} {\bibfnamefont {A.~F.}\ \bibnamefont {White}}, \bibinfo {author} {\bibfnamefont {Y.}~\bibnamefont {Gao}}, \bibinfo {author} {\bibfnamefont {A.~J.}\ \bibnamefont {Minnich}},\ and\ \bibinfo {author} {\bibfnamefont {G.~K.-L.}\ \bibnamefont {Chan}},\ }\bibfield  {title} {\bibinfo {title} {A coupled cluster framework for electrons and phonons},\ }\href {https://doi.org/10.1063/5.0033132} {\bibfield  {journal} {\bibinfo  {journal} {J. Chem. Phys.}\ }\textbf {\bibinfo {volume} {153}},\ \bibinfo {pages} {224112} (\bibinfo {year} {2020})}\BibitemShut {NoStop}%
\bibitem [{\citenamefont {Mordovina}\ \emph {et~al.}(2020)\citenamefont {Mordovina}, \citenamefont {Bungey}, \citenamefont {Appel}, \citenamefont {Knowles}, \citenamefont {Rubio},\ and\ \citenamefont {Manby}}]{PRR023262}%
  \BibitemOpen
  \bibfield  {author} {\bibinfo {author} {\bibfnamefont {U.}~\bibnamefont {Mordovina}}, \bibinfo {author} {\bibfnamefont {C.}~\bibnamefont {Bungey}}, \bibinfo {author} {\bibfnamefont {H.}~\bibnamefont {Appel}}, \bibinfo {author} {\bibfnamefont {P.~J.}\ \bibnamefont {Knowles}}, \bibinfo {author} {\bibfnamefont {A.}~\bibnamefont {Rubio}},\ and\ \bibinfo {author} {\bibfnamefont {F.~R.}\ \bibnamefont {Manby}},\ }\bibfield  {title} {\bibinfo {title} {Polaritonic coupled-cluster theory},\ }\href {https://doi.org/10.1103/PhysRevResearch.2.023262} {\bibfield  {journal} {\bibinfo  {journal} {Phys. Rev. Res.}\ }\textbf {\bibinfo {volume} {2}},\ \bibinfo {pages} {023262} (\bibinfo {year} {2020})}\BibitemShut {NoStop}%
\bibitem [{\citenamefont {Yang}\ \emph {et~al.}(2024)\citenamefont {Yang}, \citenamefont {Cui}, \citenamefont {Mahajan}, \citenamefont {Zhai}, \citenamefont {Reichman},\ and\ \citenamefont {Chan}}]{yang:2024be}%
  \BibitemOpen
  \bibfield  {author} {\bibinfo {author} {\bibfnamefont {J.}~\bibnamefont {Yang}}, \bibinfo {author} {\bibfnamefont {Z.-H.}\ \bibnamefont {Cui}}, \bibinfo {author} {\bibfnamefont {A.}~\bibnamefont {Mahajan}}, \bibinfo {author} {\bibfnamefont {H.}~\bibnamefont {Zhai}}, \bibinfo {author} {\bibfnamefont {D.~R.}\ \bibnamefont {Reichman}},\ and\ \bibinfo {author} {\bibfnamefont {G.~K.-L.}\ \bibnamefont {Chan}},\ }\href {https://arxiv.org/abs/2405.18771} {\bibinfo {title} {Benchmarking the exponential ansatz for the holstein model}} (\bibinfo {year} {2024}),\ \Eprint {https://arxiv.org/abs/2405.18771} {arXiv:2405.18771 [cond-mat.mtrl-sci]} \BibitemShut {NoStop}%
\bibitem [{\citenamefont {Shaffer}\ \emph {et~al.}(2024)\citenamefont {Shaffer}, \citenamefont {Claassen}, \citenamefont {Srivastava},\ and\ \citenamefont {Santos}}]{Shaffer:2024et}%
  \BibitemOpen
  \bibfield  {author} {\bibinfo {author} {\bibfnamefont {D.}~\bibnamefont {Shaffer}}, \bibinfo {author} {\bibfnamefont {M.}~\bibnamefont {Claassen}}, \bibinfo {author} {\bibfnamefont {A.}~\bibnamefont {Srivastava}},\ and\ \bibinfo {author} {\bibfnamefont {L.~H.}\ \bibnamefont {Santos}},\ }\bibfield  {title} {\bibinfo {title} {Entanglement and topology in su-schrieffer-heeger cavity quantum electrodynamics},\ }\href {https://doi.org/10.1103/PhysRevB.109.155160} {\bibfield  {journal} {\bibinfo  {journal} {Phys. Rev. B}\ }\textbf {\bibinfo {volume} {109}},\ \bibinfo {pages} {155160} (\bibinfo {year} {2024})}\BibitemShut {NoStop}%
\bibitem [{\citenamefont {Passetti}\ \emph {et~al.}(2023)\citenamefont {Passetti}, \citenamefont {Eckhardt}, \citenamefont {Sentef},\ and\ \citenamefont {Kennes}}]{Passetti:2023tc}%
  \BibitemOpen
  \bibfield  {author} {\bibinfo {author} {\bibfnamefont {G.}~\bibnamefont {Passetti}}, \bibinfo {author} {\bibfnamefont {C.~J.}\ \bibnamefont {Eckhardt}}, \bibinfo {author} {\bibfnamefont {M.~A.}\ \bibnamefont {Sentef}},\ and\ \bibinfo {author} {\bibfnamefont {D.~M.}\ \bibnamefont {Kennes}},\ }\bibfield  {title} {\bibinfo {title} {Cavity light-matter entanglement through quantum fluctuations},\ }\href {https://doi.org/10.1103/PhysRevLett.131.023601} {\bibfield  {journal} {\bibinfo  {journal} {Phys. Rev. Lett.}\ }\textbf {\bibinfo {volume} {131}},\ \bibinfo {pages} {023601} (\bibinfo {year} {2023})}\BibitemShut {NoStop}%
\bibitem [{\citenamefont {Weight}\ \emph {et~al.}(2024)\citenamefont {Weight}, \citenamefont {Tretiak},\ and\ \citenamefont {Zhang}}]{Weight:2024vi}%
  \BibitemOpen
  \bibfield  {author} {\bibinfo {author} {\bibfnamefont {B.~M.}\ \bibnamefont {Weight}}, \bibinfo {author} {\bibfnamefont {S.}~\bibnamefont {Tretiak}},\ and\ \bibinfo {author} {\bibfnamefont {Y.}~\bibnamefont {Zhang}},\ }\bibfield  {title} {\bibinfo {title} {Diffusion quantum monte carlo approach to the polaritonic ground state},\ }\href {https://doi.org/10.1103/PhysRevA.109.032804} {\bibfield  {journal} {\bibinfo  {journal} {Phys. Rev. A}\ }\textbf {\bibinfo {volume} {109}},\ \bibinfo {pages} {032804} (\bibinfo {year} {2024})}\BibitemShut {NoStop}%
\bibitem [{\citenamefont {Weight}\ and\ \citenamefont {Zhang}()}]{Weight:2024af}%
  \BibitemOpen
  \bibfield  {author} {\bibinfo {author} {\bibfnamefont {B.~M.}\ \bibnamefont {Weight}}\ and\ \bibinfo {author} {\bibfnamefont {Y.}~\bibnamefont {Zhang}},\ }\bibfield  {title} {\bibinfo {title} {Auxiliary field quantum monte carlo for electron-photon correlation},\ }\href@noop {} {\bibinfo  {journal} {In preparation}\ }\BibitemShut {NoStop}%
\bibitem [{\citenamefont {Feinberg}\ \emph {et~al.}(1990)\citenamefont {Feinberg}, \citenamefont {Ciuchi},\ and\ \citenamefont {de~Pasquale}}]{Feinberg:1990sp}%
  \BibitemOpen
\bibfield  {journal} {  }\bibfield  {author} {\bibinfo {author} {\bibfnamefont {D.}~\bibnamefont {Feinberg}}, \bibinfo {author} {\bibfnamefont {S.}~\bibnamefont {Ciuchi}},\ and\ \bibinfo {author} {\bibfnamefont {F.}~\bibnamefont {de~Pasquale}},\ }\bibfield  {title} {\bibinfo {title} {Squeezing phenomena in interacting electron-phonon systems},\ }\href {https://doi.org/10.1142/S0217979290000656} {\bibfield  {journal} {\bibinfo  {journal} {International Journal of Modern Physics B}\ }\textbf {\bibinfo {volume} {04}},\ \bibinfo {pages} {1317} (\bibinfo {year} {1990})},\ \bibinfo {note} {doi: 10.1142/S0217979290000656}\BibitemShut {NoStop}%
\bibitem [{\citenamefont {Trapper}\ \emph {et~al.}(1994)\citenamefont {Trapper}, \citenamefont {Fehske}, \citenamefont {Deeg},\ and\ \citenamefont {B{\"u}ttner}}]{Trapper:1994tg}%
  \BibitemOpen
  \bibfield  {author} {\bibinfo {author} {\bibfnamefont {U.}~\bibnamefont {Trapper}}, \bibinfo {author} {\bibfnamefont {H.}~\bibnamefont {Fehske}}, \bibinfo {author} {\bibfnamefont {M.}~\bibnamefont {Deeg}},\ and\ \bibinfo {author} {\bibfnamefont {H.}~\bibnamefont {B{\"u}ttner}},\ }\bibfield  {title} {\bibinfo {title} {Electron correlations and quantum lattice vibrations in strongly coupled electron-phonon systems: A variational slave boson approach},\ }\href {https://doi.org/10.1007/BF01314251} {\bibfield  {journal} {\bibinfo  {journal} {Zeitschrift f{\"u}r Physik B Condensed Matter}\ }\textbf {\bibinfo {volume} {93}},\ \bibinfo {pages} {465} (\bibinfo {year} {1994})}\BibitemShut {NoStop}%
\bibitem [{\citenamefont {Barone}\ \emph {et~al.}(2008)\citenamefont {Barone}, \citenamefont {Raimondi}, \citenamefont {Capone}, \citenamefont {Castellani},\ and\ \citenamefont {Fabrizio}}]{Barone:2008ug}%
  \BibitemOpen
  \bibfield  {author} {\bibinfo {author} {\bibfnamefont {P.}~\bibnamefont {Barone}}, \bibinfo {author} {\bibfnamefont {R.}~\bibnamefont {Raimondi}}, \bibinfo {author} {\bibfnamefont {M.}~\bibnamefont {Capone}}, \bibinfo {author} {\bibfnamefont {C.}~\bibnamefont {Castellani}},\ and\ \bibinfo {author} {\bibfnamefont {M.}~\bibnamefont {Fabrizio}},\ }\bibfield  {title} {\bibinfo {title} {Gutzwiller scheme for electrons and phonons: The half-filled hubbard-holstein model},\ }\href {https://doi.org/10.1103/PhysRevB.77.235115} {\bibfield  {journal} {\bibinfo  {journal} {Phys. Rev. B}\ }\textbf {\bibinfo {volume} {77}},\ \bibinfo {pages} {235115} (\bibinfo {year} {2008})}\BibitemShut {NoStop}%
\bibitem [{\citenamefont {Walls}(1983)}]{Walls:1983tv}%
  \BibitemOpen
  \bibfield  {author} {\bibinfo {author} {\bibfnamefont {D.~F.}\ \bibnamefont {Walls}},\ }\bibfield  {title} {\bibinfo {title} {Squeezed states of light},\ }\href {https://doi.org/10.1038/306141a0} {\bibfield  {journal} {\bibinfo  {journal} {Nature}\ }\textbf {\bibinfo {volume} {306}},\ \bibinfo {pages} {141} (\bibinfo {year} {1983})}\BibitemShut {NoStop}%
\bibitem [{\citenamefont {Cohen-Tannoudji}\ \emph {et~al.}(1997)\citenamefont {Cohen-Tannoudji}, \citenamefont {Dupont-Roc},\ and\ \citenamefont {Grynberg}}]{CohenTannoudji1997}%
  \BibitemOpen
  \bibfield  {author} {\bibinfo {author} {\bibfnamefont {C.}~\bibnamefont {Cohen-Tannoudji}}, \bibinfo {author} {\bibfnamefont {J.}~\bibnamefont {Dupont-Roc}},\ and\ \bibinfo {author} {\bibfnamefont {G.}~\bibnamefont {Grynberg}},\ }\href {https://www.ebook.de/de/product/3737960/claude_cohen_tannoudji_jacques_dupont_roc_gilbert_grynberg_photons_and_atoms_introduction_to_quantum_electrodynamics.html} {\emph {\bibinfo {title} {Photons and Atoms: Introduction to Quantum Electrodynamics}}}\ (\bibinfo  {publisher} {VCH PUBN},\ \bibinfo {year} {1997})\BibitemShut {NoStop}%
\bibitem [{\citenamefont {Rokaj}\ \emph {et~al.}(2018)\citenamefont {Rokaj}, \citenamefont {Welakuh}, \citenamefont {Ruggenthaler},\ and\ \citenamefont {Rubio}}]{Rokaj2018JPBAMOP}%
  \BibitemOpen
  \bibfield  {author} {\bibinfo {author} {\bibfnamefont {V.}~\bibnamefont {Rokaj}}, \bibinfo {author} {\bibfnamefont {D.~M.}\ \bibnamefont {Welakuh}}, \bibinfo {author} {\bibfnamefont {M.}~\bibnamefont {Ruggenthaler}},\ and\ \bibinfo {author} {\bibfnamefont {A.}~\bibnamefont {Rubio}},\ }\bibfield  {title} {\bibinfo {title} {Light{\textendash}matter interaction in the long-wavelength limit: no ground-state without dipole self-energy},\ }\href {https://doi.org/10.1088/1361-6455/aa9c99} {\bibfield  {journal} {\bibinfo  {journal} {J. Phys. B: At. Mol. Opt. Phys.}\ }\textbf {\bibinfo {volume} {51}},\ \bibinfo {pages} {034005} (\bibinfo {year} {2018})}\BibitemShut {NoStop}%
\bibitem [{\citenamefont {Bailes}\ \emph {et~al.}(2021)\citenamefont {Bailes}, \citenamefont {Berger}, \citenamefont {Brady}, \citenamefont {Branchesi}, \citenamefont {Danzmann}, \citenamefont {Evans}, \citenamefont {Holley-Bockelmann}, \citenamefont {Iyer}, \citenamefont {Kajita}, \citenamefont {Katsanevas}, \citenamefont {Kramer}, \citenamefont {Lazzarini}, \citenamefont {Lehner}, \citenamefont {Losurdo}, \citenamefont {L{\"u}ck}, \citenamefont {McClelland}, \citenamefont {McLaughlin}, \citenamefont {Punturo}, \citenamefont {Ransom}, \citenamefont {Raychaudhury}, \citenamefont {Reitze}, \citenamefont {Ricci}, \citenamefont {Rowan}, \citenamefont {Saito}, \citenamefont {Sanders}, \citenamefont {Sathyaprakash}, \citenamefont {Schutz}, \citenamefont {Sesana}, \citenamefont {Shinkai}, \citenamefont {Siemens}, \citenamefont {Shoemaker}, \citenamefont {Thorpe}, \citenamefont {van~den Brand},\ and\ \citenamefont {Vitale}}]{Bailes:2021tz}%
  \BibitemOpen
  \bibfield  {author} {\bibinfo {author} {\bibfnamefont {M.}~\bibnamefont {Bailes}}, \bibinfo {author} {\bibfnamefont {B.~K.}\ \bibnamefont {Berger}}, \bibinfo {author} {\bibfnamefont {P.~R.}\ \bibnamefont {Brady}}, \bibinfo {author} {\bibfnamefont {M.}~\bibnamefont {Branchesi}}, \bibinfo {author} {\bibfnamefont {K.}~\bibnamefont {Danzmann}}, \bibinfo {author} {\bibfnamefont {M.}~\bibnamefont {Evans}}, \bibinfo {author} {\bibfnamefont {K.}~\bibnamefont {Holley-Bockelmann}}, \bibinfo {author} {\bibfnamefont {B.~R.}\ \bibnamefont {Iyer}}, \bibinfo {author} {\bibfnamefont {T.}~\bibnamefont {Kajita}}, \bibinfo {author} {\bibfnamefont {S.}~\bibnamefont {Katsanevas}}, \bibinfo {author} {\bibfnamefont {M.}~\bibnamefont {Kramer}}, \bibinfo {author} {\bibfnamefont {A.}~\bibnamefont {Lazzarini}}, \bibinfo {author} {\bibfnamefont {L.}~\bibnamefont {Lehner}}, \bibinfo {author} {\bibfnamefont {G.}~\bibnamefont {Losurdo}}, \bibinfo {author} {\bibfnamefont {H.}~\bibnamefont {L{\"u}ck}}, \bibinfo {author} {\bibfnamefont
  {D.~E.}\ \bibnamefont {McClelland}}, \bibinfo {author} {\bibfnamefont {M.~A.}\ \bibnamefont {McLaughlin}}, \bibinfo {author} {\bibfnamefont {M.}~\bibnamefont {Punturo}}, \bibinfo {author} {\bibfnamefont {S.}~\bibnamefont {Ransom}}, \bibinfo {author} {\bibfnamefont {S.}~\bibnamefont {Raychaudhury}}, \bibinfo {author} {\bibfnamefont {D.~H.}\ \bibnamefont {Reitze}}, \bibinfo {author} {\bibfnamefont {F.}~\bibnamefont {Ricci}}, \bibinfo {author} {\bibfnamefont {S.}~\bibnamefont {Rowan}}, \bibinfo {author} {\bibfnamefont {Y.}~\bibnamefont {Saito}}, \bibinfo {author} {\bibfnamefont {G.~H.}\ \bibnamefont {Sanders}}, \bibinfo {author} {\bibfnamefont {B.~S.}\ \bibnamefont {Sathyaprakash}}, \bibinfo {author} {\bibfnamefont {B.~F.}\ \bibnamefont {Schutz}}, \bibinfo {author} {\bibfnamefont {A.}~\bibnamefont {Sesana}}, \bibinfo {author} {\bibfnamefont {H.}~\bibnamefont {Shinkai}}, \bibinfo {author} {\bibfnamefont {X.}~\bibnamefont {Siemens}}, \bibinfo {author} {\bibfnamefont {D.~H.}\ \bibnamefont {Shoemaker}}, \bibinfo
  {author} {\bibfnamefont {J.}~\bibnamefont {Thorpe}}, \bibinfo {author} {\bibfnamefont {J.~F.~J.}\ \bibnamefont {van~den Brand}},\ and\ \bibinfo {author} {\bibfnamefont {S.}~\bibnamefont {Vitale}},\ }\bibfield  {title} {\bibinfo {title} {Gravitational-wave physics and astronomy in the 2020s and 2030s},\ }\href {https://doi.org/10.1038/s42254-021-00303-8} {\bibfield  {journal} {\bibinfo  {journal} {Nat. Rev. Phys.}\ }\textbf {\bibinfo {volume} {3}},\ \bibinfo {pages} {344} (\bibinfo {year} {2021})}\BibitemShut {NoStop}%
\bibitem [{\citenamefont {Zhang}(2023)}]{openms2023}%
  \BibitemOpen
  \bibfield  {author} {\bibinfo {author} {\bibfnamefont {Y.}~\bibnamefont {Zhang}},\ }\href {https://doi.org/10.11578/dc.20230602.3} {\bibinfo {title} {Multiscale ecosystem for solving maxwell-schrodinger equations of open quantum systems (openms)}} (\bibinfo {year} {2023})\BibitemShut {NoStop}%
\bibitem [{\citenamefont {Hauschild}\ and\ \citenamefont {Pollmann}(2018)}]{tenpy}%
  \BibitemOpen
  \bibfield  {author} {\bibinfo {author} {\bibfnamefont {J.}~\bibnamefont {Hauschild}}\ and\ \bibinfo {author} {\bibfnamefont {F.}~\bibnamefont {Pollmann}},\ }\bibfield  {title} {\bibinfo {title} {{Efficient numerical simulations with Tensor Networks: Tensor Network Python (TeNPy)}},\ }\href {https://doi.org/10.21468/SciPostPhysLectNotes.5} {\bibfield  {journal} {\bibinfo  {journal} {SciPost Phys. Lect. Notes}\ ,\ \bibinfo {pages} {5}} (\bibinfo {year} {2018})},\ \bibinfo {note} {code available from \url{https://github.com/tenpy/tenpy}},\ \Eprint {https://arxiv.org/abs/1805.00055} {arXiv:1805.00055} \BibitemShut {NoStop}%
\bibitem [{\citenamefont {Weber}\ \emph {et~al.}(2024)\citenamefont {Weber}, \citenamefont {dos Anjos~Cunha}, \citenamefont {Morales}, \citenamefont {Rubio},\ and\ \citenamefont {Zhang}}]{Weber:2024af}%
  \BibitemOpen
  \bibfield  {author} {\bibinfo {author} {\bibfnamefont {L.}~\bibnamefont {Weber}}, \bibinfo {author} {\bibfnamefont {L.}~\bibnamefont {dos Anjos~Cunha}}, \bibinfo {author} {\bibfnamefont {M.~A.}\ \bibnamefont {Morales}}, \bibinfo {author} {\bibfnamefont {A.}~\bibnamefont {Rubio}},\ and\ \bibinfo {author} {\bibfnamefont {S.}~\bibnamefont {Zhang}},\ }\bibfield  {title} {\bibinfo {title} {Phaseless auxiliary-field quantum monte carlo method for cavity-qed matter systems},\ }\href {https://doi.org/10.48550/arXiv.2410.18838} {\bibfield  {journal} {\bibinfo  {journal} {arXiv preprint arXiv:2410.18838}\ } (\bibinfo {year} {2024})}\BibitemShut {NoStop}%
\bibitem [{\citenamefont {Zhang}\ \emph {et~al.}(2015)\citenamefont {Zhang}, \citenamefont {Yam}, \citenamefont {Kwok},\ and\ \citenamefont {Chen}}]{Zhang:2015ua}%
  \BibitemOpen
  \bibfield  {author} {\bibinfo {author} {\bibfnamefont {Y.}~\bibnamefont {Zhang}}, \bibinfo {author} {\bibfnamefont {C.}~\bibnamefont {Yam}}, \bibinfo {author} {\bibfnamefont {Y.}~\bibnamefont {Kwok}},\ and\ \bibinfo {author} {\bibfnamefont {G.}~\bibnamefont {Chen}},\ }\bibfield  {title} {\bibinfo {title} {A variational approach for dissipative quantum transport in a wide parameter space},\ }\href {https://doi.org/10.1063/1.4930847} {\bibfield  {journal} {\bibinfo  {journal} {J. Chem. Phys.}\ }\textbf {\bibinfo {volume} {143}},\ \bibinfo {pages} {104112} (\bibinfo {year} {2015})}\BibitemShut {NoStop}%
\end{thebibliography}%

\appendix
\clearpage
\begin{widetext}
\begin{center}
  {\Large\bf Supplementary Materials for ``Light-Matter Hybridization and Entanglement from First-Principles"}
\end{center}

This appendix provides detailed derivations of the generalized QEDHF theories with various squeeze ans{\"a}tze.

\section{Properties of displacement and squeeze operators}
\label{sec:app:operator_prop}
Before deriving the detailed equations, we discuss the properties and physical meanings of the displacement and squeeze operators.

\subsection{Squeeze operator}
The squeeze operator is defined as
\begin{align}
    \label{eq:squeeze_op}
    \hat{S} ( F_{\alpha} ) = \prod_{\alpha} \text{exp}
    \left[ \frac{1}{2}
           \left(F^*_{\alpha} \hat{b}^2_\alpha -
           F_{\alpha}\hat{b}^{\dag 2}_\alpha \right) \right],
\end{align}
where $F_{\alpha} \equiv r_{\alpha}e^{i\theta_\alpha}$ is generally complex. The squeeze operator is a unitary operator, satisfying $\hat{S}(F)\hat{S}^\dag(F)=\hat{S}^\dag(F)\hat{S}(F)=\hat{I}$,
where $\hat{I}$ is the identity operator. The squeeze operator can be applied to any state.
For instance, its action on the vacuum (or coherent) state produces a squeezed (or squeezed coherent) state. Squeezed states are widely utilized in precision measurements, such as interferometric measurements in the Laser Interferometer Gravitational-Wave Observatory (LIGO)~\cite{Bailes:2021tz}.

Due to the presence of ladder operators within the exponential, the action of the squeeze operator on individual bosonic creation or annihilation operators introduces a mixture of both creation and annihilation operators,
\begin{align}
    \hat{S}^{\dag} (F_{\alpha}) \hat{b}^{\dag}_{\alpha} \hat{S} (F_{\alpha})
        &= \cosh(r_{\alpha}) \hat{b}^{\dag}_{\alpha} - \sinh(r_{\alpha}) \hat{b}_{\alpha}, \\
    \hat{S}^{\dag} (F_{\alpha}) \hat{b}_{\alpha} \hat{S} (F_{\alpha})
        &= \cosh(r_{\alpha}) \hat{b}_{\alpha} - \sinh(r_{\alpha}) \hat{b}^{\dag}_{\alpha}.
\end{align}

\subsection{Displacement operator}

The displacement operator in quantum optics is defined as
\begin{align}
    \label{eq:cs_op}
    \hat{D} ( z_{\alpha} ) = \prod_{\alpha} \text{exp}
    \left[
    - \left(z^*_{\alpha} \hat{b}_{\alpha} - z_{\alpha}\hat{b}^{\dag}_{\alpha} \right) \right].
\end{align}
which displaces a localized state in phase space by a magnitude of $z_\alpha$.
Specifically, when $z_\alpha$ is a scalar, the action of the displacement operator on the vacuum state produces a coherent state $\ket{z_\alpha}$.
The bosonic operators in the coherent state representation become,
\begin{align}\label{eq:displacez}
    \hat{D}^{\dag} (z_{\alpha}) \hat{b}^{\dag}_{\alpha} \hat{D} (z_{\alpha})
        &= \hat{b}^{\dag}_{\alpha} - z^*_{\alpha}, \\
    \hat{D}^{\dag} (z_{\alpha}) \hat{b}_{\alpha} \hat{D} (z_{\alpha})
        &= \hat{b}_{\alpha} - z_{\alpha}.
\end{align}

In the study of electron-boson interactions, it is often beneficial to introduce an electronic operator $\hat{f}$ in the displacement operator to account for orbital relaxation effects.
The generalized displacement operator is then defined as:
\begin{align}
    \label{eq:vt_op}
    \hat{D} ( \hat{f}_{\alpha} ) = \prod_{\alpha} \text{exp}
    \left[
    -\hat{f}_{\alpha} \left( \hat{b}_{\alpha} - \hat{b}^{\dag}_{\alpha} \right) \right].
\end{align}

In principle, there are no restrictions on the choice of the $\hat{f}$ operator. For an electron-boson interacting system with an interaction Hamiltonian of the form $\hat{H}_{\text{int}} = \sum_\alpha \hat{O}^\alpha_f \hat{O}^\alpha_b$, where $\hat{O}_f$ and $\hat{O}_b$ are fermionic and bosonic operators respectively, a natural choice is $\hat{f} = \hat{O}_f$. This choice partially decouples the electron-boson interaction through the (variational) Lang-Firsov transformation~\cite{Zhang:2023vt, Zhang:2015ua}. For light-matter interactions considered in this work, $\hat{f}_\alpha = \frac{f_\alpha \boldsymbol{\lambda}_\alpha \cdot \boldsymbol{D}}{\sqrt{2\omega_\alpha}}$, where $f_\alpha$ is a variational parameter controlling the strength of displacement.
Similar to the coherent state displacement, the similarity transformation with $\hat{D}(\hat{f})$ modifies the bosonic operator as
\[
  \hat{D}^\dag(\hat{f}_\alpha) b^\dag_\alpha \hat{D}(z) = b^\dag - \hat{f}_\alpha.
\]

If the orbital relaxation effects are neglected and the electronic DOF in $f_\alpha$ is traced out, $\hat{D}(\hat{f})$ reduces to the coherent state displacement $\hat{D}(z_\alpha)$. However, the introduction of the electronic operator $\hat{f}$ grants $\hat{D}(\hat{f})$ additional variational flexibility. Notably, $\hat{D}(\hat{f})$ introduces orbital-dependent displacement to the electronic operator, whereas the coherent state displacement $\hat{D}(z_\alpha)$ does not affect electronic operators.

\textit{In this work, we focus on applications in molecular systems with a real-Gaussian basis set. Therefore, both $F$ and $z$ are treated as real parameters. For simplicity, we omit complex notation in the following discussion, although the extension to complex cases is straightforward.}

\section{Wavefunction ans\"atze for QED problems}

As discussed in the main text, any photonic representation can be expressed as $\hat{U}\ket{0}$. The corresponding wavefunction ansatz for coupled electron-boson problems becomes
\begin{equation}
    \ket{\Theta} = \sum_n C_n \hat{U}\ket{\Phi}\otimes\ket{n_p}.
\end{equation}
Thus, the ground state energy can be obtained by solving $\hat{H}\ket{\Psi} = E\ket{\Psi}$, which is equivalent to
\begin{equation}
    \hat{U}^\dag \hat{H} \hat{U} \ket{\Phi}\otimes\ket{0_p} = E \ket{\Phi}\otimes\ket{0_p}.
\end{equation}
After integrating over the photonic DOFs, it reduces to
\begin{equation}
     \sum_n |C_n|^2\bra{n_p} \hat{U}^\dag \hat{H} \hat{U} \ket{n_p} \ket{\Phi} = E\ket{\Phi}.
\end{equation}
Here, we define an effective Hamiltonian as $\hat{H}^e_{\text{eff}} =\hat{U}^\dag \hat{H} \hat{U}$.
Once the effective Hamiltonian is known, and after integrating the photonic DOFs, the molecular QED problems can be solved in the same way as pure electronic problems.
In the mean-field approximation, where the electron-photon wavefunction is decomposed as a single product, the QED mean-field solution becomes
\[
\bra{0_p}\hat{U}^\dag \hat{H} \hat{U} \ket{0_p} \ket{\text{HF}} = E\ket{\text{HF}}.
\]
Once the one-body and two-body integrals are obtained under a specific photonic representation, the QEDHF energies can be readily calculated using standard HF procedures.

\section{QEDHF formalisms with joint transformations}

Next, we derive the effective Hamiltonian within different wavefunction ans\"atze. As discussed in the main text, there are four combinations of displacement and squeeze operators since $\hat{S}$ and $\hat{D}$ operators do not commute. First, we demonstrate in Sec.~\ref{sec:relation} that $\hat{S}(F)\hat{D}$ is related to $\hat{D}\hat{S}(F)$ via a rotation. Second, in Sec.~\ref{sec:singletransform}, we provide the Hamiltonian after the similarity transformation with a single operator (either displacement or squeezing). Finally, in Secs.~\ref{sec:qedhfscs}--\ref{sec:qedhfSGS}, we derive the similarity-transformed Hamiltonian for joint displacement and squeeze operators across the four combinations.

\subsection{Relation between $\hat{D}(z)\hat{S}$ and $\hat{S}\hat{D}(z)$ similarity transformations}
\label{sec:relation}

Although $\hat{S}\hat{D} \neq \hat{D}\hat{S}$, they are related via a rotation,
\begin{align}
(\hat{S}\hat{D})^\dag \hat{b}_\alpha \hat{S}\hat{D} = & \hat{D}^\dag [\cosh(r)\hat{b} - \sinh(r)\hat{b}^\dag]\hat{D} =
\cosh(r) [\hat{b} - z_\alpha] - \sinh(r)[\hat{b}^\dag - z^*]
\nonumber\\
= & \cosh(r)\hat{b} - \sinh(r) \hat{b}^\dag - ze^{-r}s
\nonumber\\
= & \hat{S}^\dag[\hat{D}^\dag(ze^{-r}) \hat{b} \hat{D}(ze^{-r})]\hat{S}(r).
\end{align}
Thus, a unitary transformation with $\hat{S}(r)\hat{D}(z)$ is equivalent to $\hat{D}(\tilde{z})\hat{S}(r)$, where the effective displacement is $\tilde{z} = e^{-r}z$.

However, if the displacement scalar is replaced by an electronic operator, $\hat{f}$, the effects of $\hat{S}(r)\hat{D}(\hat{f})$ and $\hat{D}(\hat{f})\hat{S}(r)$ are not equivalent. The action of $\hat{S}(r)\hat{D}(\hat{f})$ and $\hat{D}(\hat{f})\hat{S}(r)$ on the bosonic operator can still be equivalent, because
\begin{align}
[\hat{S}\hat{D}(\hat{f})]^\dag \hat{b}_\alpha \hat{S}\hat{D}(\hat{f}) = & \hat{D}^\dag [\cosh(r)\hat{b} - \sinh(r)\hat{b}^\dag]\hat{D} =
\cosh(r) [\hat{b} - \hat{f}_\alpha] - \sinh(r)[\hat{b}^\dag - \hat{f}^\dag_\alpha]
\nonumber\\
= & \cosh(r)\hat{b} - \sinh(r) \hat{b}^\dag - [\cosh(r)\hat{f} - \hat{f}^\dag\sinh(r)]
\nonumber\\
= & \hat{S}^\dag\hat{D}^\dag(\hat{\mathcal{F}}) \hat{b} \hat{D}(\hat{\mathcal{F}})\hat{S}(r),
\end{align}
where $\hat{\mathcal{F}}$ is the effective displacement.

However, due to the use of the fermionic operator in the displacement, the action of $\hat{S}(r)\hat{D}(\hat{f})$ and $\hat{D}(\hat{f})\hat{S}(r)$ on the fermionic operator is no longer equivalent. Specifically
\[
[\hat{S}\hat{D}(\hat{f})]^\dag \hat{c} \hat{S}\hat{D}(\hat{f}) = \hat{D}^\dag(\hat{f}) \hat{c} \hat{D}(\hat{f}),
\]
where the squeeze operator does not affect the electronic operator, as fermionic and bosonic operators commute. In contrast
\[
 [\hat{D}\hat{S}(r)]^\dag\hat{c} [\hat{D}\hat{S}(r)] = \hat{S}^\dag(r)\left[\hat{c}\hat{\mathcal{X}} \right]\hat{S}(r),
\]
where both squeezing and displacement contribute. Consequently, the unitary transformations with $\hat{S}\hat{D}(\hat{f})$ and $\hat{D}(\hat{f})\hat{S}(r)$ are not equivalent. Only the latter introduces squeezing-induced renormalization effects on the electronic operator, leading to profound impacts in the variational transformation. This is the primary reason why the latter is discussed in the main text.

\subsection{Hamiltonians with single transformation}
\label{sec:singletransform}

Here, we derive the effective Hamiltonian for three types of single transformations: coherent state, variational displacement, and variational squeeze transformations.

\noindent
{\bf a. Coherent state.}
Using the properties of displacement operators in Eq.~\ref{eq:displacez}, the Hamiltonian in the coherent state representation can be straightforwardly derived as~\cite{haugland_coupled_2020},
\begin{equation}\label{eq:hamcs}
  \hat{\mathcal{H}}_{D} =
  \hat{H}_e
  + \sqrt{\frac{\omega_\alpha}{2}}[\boldsymbol{\lambda}_\alpha\cdot(\boldsymbol{D}-\langle\boldsymbol{D})\rangle](\hb^\dag_\alpha+\hb_\alpha)
  + \frac{1}{2}[\boldsymbol{\lambda}_\alpha\cdot(\boldsymbol{D}-\langle\boldsymbol{D}\rangle)]^2
  + \sum_{\alpha} \omega_{\alpha}(\hb^\dag_\alpha \hb_\alpha+\frac{1}{2}).
\end{equation}

\noindent
{\bf b. Variational displacement transformation.}
The Hamiltonian under the Variational displacement transformation is~\cite{Zhang:2023vt},
\begin{equation}\label{eq:hamvlf}
  \hat{\mathcal{H}}_{VT} = \hat{\mathcal{H}}_e({X})
  + \sum_{\alpha}\sqrt{\frac{\omega_\alpha}{2}}(\Delta\lambda_\alpha) \mathbf{e}_\alpha\cdot\mathbf{D}(\hb^\dag_\alpha+\hb_\alpha )
  + \frac{(\Delta\lambda_\alpha)^2}{2}(\mathbf{e}_\alpha\cdot\mathbf{D})^2
  + \sum_{\alpha} \omega_{\alpha}(\hb^\dag_\alpha \hb_\alpha+\frac{1}{2}).
\end{equation}
Where $\hat{\mathcal{H}}_e(\hat{X})$ is the dressed electronic Hamiltonian,
\begin{equation}
  \hat{\mathcal{H}}_e({\hat{X}}) = \sum_{ij} h_{ij}\hat{X}\hat{X} + \frac{1}{2}\sum_{ijkl} I_{ijkl} \hat{X}^\dag \hat{X}^\dag \hat{X} \hat{X}_l.
\end{equation}
Here, $\hat{X} = \exp\left[-f_\alpha \frac{\ldotd}{\sqrt{2\omega_\alpha}} (\hb^\dag_\alpha - \hb_\alpha)\right]$ is the displacement operator acting on the electronic DOFs, introducing orbital-dependent displacements on the electronic operators. The dressed electronic Hamiltonian reduces to the original (polaron) Hamiltonian when $f_\alpha = 0$ ($f_\alpha =1$).

\noindent
{\bf c. Variational Squeeze transformation.}
The effective Hamiltonian after the similarity transformation with the squeeze operator becomes,
\begin{equation}\label{eq:hams}
  \hat{\mathcal{H}}_S = \hat{H}_e + \frac{1}{2}(\ldotd)^2 + \sum_\alpha \sqrt{\frac{\omega_\alpha}{2}}(\ldotd)e^{-r}(\hb^\dag_\alpha + \hb_\alpha) +
  \sum_\alpha\omega_\alpha \left\{\cosh(2r) (\hb^\dag_\alpha \hb_\alpha+\frac{1}{2}) -\sinh(2r)(\hb^{\dag 2}_\alpha + \hb^2_\alpha) \right\}.
\end{equation}
In contrast to the Variational displacement and coherent state transformations, the photonic Hamiltonian is formally altered by the squeezing transformation. However, since $\cosh(2r) \geq 1$, the expectation value of the photonic Hamiltonian still has a lower bound of $\frac{\omega_\alpha}{2}$. Thus, the squeeze operator does not change the photon ground state energy.


\subsection{QEDHF formalism with the Variational Gaussian Squeezed State (GSS)}\label{sec:qedhfGSS}

This section presents the detailed derivation of the QEDHF formalism within the Gaussian Squeezed State (GSS), where the wavefunction ansatz is defined as
\begin{equation}
    \hat{U}(\hat{f}_\alpha, F) = \hat{D}(\hat{f}_\alpha) \hat{S}(F).
\end{equation}
The action of this transformation on bosonic and electronic operators is as follows
\begin{align}
    &\hat{\mathcal{U}}^\dag \hat{b}^\dag \hat{\mathcal{U}} = \hat{S}^\dag(r)[\hat{b}^\dag - z_\alpha] \hat{S}(r)
    = \cosh(r) \hat{b}^\dag - \sinh(r) \hat{b} - z_\alpha.
    \\
    &\hat{\mathcal{U}}^\dag [\hat{b}^\dag + \hat{b}] \hat{\mathcal{U}} = \hat{S}^\dag(r)[\hat{b}^\dag + \hat{b} - z_\alpha] \hat{S}(r),
\end{align}
and
\begin{align}
    \hat{\mathcal{U}}^\dag \hat{c}_i \hat{\mathcal{U}}
    = \sum_j \hat{c}_j \hat{\mathcal{X}}_{ij},
\end{align}
where $\hat{\mathcal{X}}_{ij}$ is given by
\begin{align}
\hat{\mathcal{X}}_{ij} =& \hat{S}^\dag(r)\left[
    e^{-\frac{f_\alpha\boldsymbol{\lambda}\cdot\boldsymbol{D}}{\sqrt{2\omega}}\left(\hb^\dag_\alpha - \hb_\alpha\right)}
    \Big|_{ij}\right]\hat{S}(r)
    \nonumber\\
    =& \exp\left[-\frac{f_\alpha\boldsymbol{\lambda}\cdot\boldsymbol{D}}{\sqrt{2\omega}}\left(\hb^\dag_\alpha - \hb_\alpha\right)\left[\cosh(r)+\sinh(r)\right)
    \right]\Big|_{ij}
    \nonumber\\
    =& \exp\left[-\frac{f_\alpha\boldsymbol{\lambda}\cdot\boldsymbol{D}}{\sqrt{2\omega}}\left(\hb^\dag_\alpha - \hb_\alpha\right)e^{r_\alpha}
    \right]\Big|_{ij}.
\end{align}

The Hamiltonian after the similarity transformation with the unitary operator $\hat{U}(\hat{f}_\alpha, F)$ becomes
\begin{equation}
  \hat{\mathcal{H}}_{GSS} = \hat{S}^\dag(r)\left[\hat{D}^\dag(f_\alpha)\hat{H}\hat{D}(f_\alpha)\right]\hat{S}(r)
  \equiv
  \hat{S}^\dag(r)\hat{\mathcal{H}}_{VT}\hat{S}(r),
\end{equation}
where $\hat{\mathcal{H}}_{VT}$ is the Hamiltonian from the VT transformation, as shown in Eq.~\ref{eq:hamvlf}.

The total Hamiltonian after the GSS transformation becomes,
\begin{align}\label{appeq:hcs}
\hat{\mathcal{H}}_{GSS} =&
 \hat{\mathcal{H}}_e(\mathcal{X})
 + \sum_{\alpha} e^{-r}\sqrt{\frac{\omega_\alpha}{2}}(\Delta\lambda_\alpha) \mathbf{e}_\alpha\cdot\mathbf{D}(\hb^\dag_\alpha+\hb_\alpha )
 + \frac{(\Delta\lambda_\alpha)^2}{2}(\mathbf{e}_\alpha\cdot\mathbf{D})^2
  \nonumber\\ &
 + \sum_\alpha\omega_\alpha \left\{\cosh(2r) (\hb^\dag_\alpha \hb_\alpha + \frac{1}{2}) -\sinh(2r)(\hb^{\dag 2}_\alpha + \hb^2_\alpha) \right\}
\end{align}
where $\hat{\mathcal{H}}_e(\hat{\mathcal{X}})$ is defined as
\[
\hat{\mathcal{H}}_e(\hat{\mathcal{X}}) = \hat{S}^\dag(F) \hat{\mathcal{H}}_e(\hat{X})\hat{S}(F).
\]
which is formally the same as $\hat{\mathcal{H}}_e(\hat{X})$ with $\hat{X}$ replaced by $\hat{\mathcal{X}}$. The latter is
\[
  \hat{\mathcal{X}} = \exp\left[-f_\alpha e^{r}\frac{\ldothatd}{\sqrt{2\omega_\alpha}}(\hb^\dag_\alpha - \hb_\alpha)\right].
\]

Like the VT-QEDHF case, the $\hat{\mathcal{X}}$ operators introduce the Franck-Condon factors to the one-body and two-body integrals. Suppose the eigenvalues of $\ldotd$ are $\{\eta_p\}$. The Gaussian factors for the one-body integrals are
\begin{align}
    G^\alpha_{pq} = \exp\left[\frac{-f^2_\alpha (\eta_p - \eta_q)^2 e^{2r_\alpha}}{4\omega^2_\alpha}\right]
\end{align}
The gradient of $G^\alpha_{pq}$ with respect to $r_\alpha$ is:
\begin{equation}
  \frac{\partial G^\alpha}{\partial r_\alpha} = \left[ \frac{-f^2_\alpha (\eta_p - \eta_q)^2 e^{2r_\alpha}}{2\omega^2_\alpha}\right]G^\alpha
\end{equation}

However, such calculations require diagonalizing $\ldotd$ at every SCF iteration and evaluating the factors in the molecular orbital (MO) representation. To avoid the overhead of diagonalization, we first transform the Hamiltonian into the eigenspace of $\bra{\mu}\boldsymbol{\lambda}\cdot\boldsymbol{D}\ket{\nu}$ (the dipole basis set) and introduce $\{\eta_p\}$ as additional variational variables. The original PF Hamiltonian in the dipole basis set becomes,
\begin{align}
    \hat{H}^d = \sum_{pq} h_{pq} \hat{c}^\dag_p \hat{c}_q + \frac{1}{2}\sum_{pqrs} I_{pqrs} \hat{c}^\dag_p \hat{c}^\dag_q \hat{c}_r \hat{c}_s + \sum_{pq}\sqrt{\frac{\omega_\alpha}{2}}g_{p} \hat{c}^\dag_p \hat{c}_p (\hat{b}^\dag_\alpha + \hat{b}_\alpha) + \frac{1}{2} [\sum_p g_p \hat{c}^\dag_p \hat{c}_p]^2
\end{align}
where $\{g_p\}$ are the eigenvalues of the dipole coupling matrix in the AO basis ($\bra{\mu}\boldsymbol{\lambda}\cdot\boldsymbol{D}\ket{\nu}$). The electronic displacement operator $\hat{\mathcal{X}}$ becomes
$\hat{\mathcal{X}} = \exp\left[-f_\alpha e^{r}\frac{\eta_p \hat{c}^\dag_p \hat{c}_p}{\sqrt{2\omega_\alpha}}(\hb^\dag_\alpha - \hb_\alpha)\right]$.
Consequently, the energy functional in the dipole basis set becomes:
\begin{align}
  E = & \sum_{pq} h_{pq} \rho_{pq} G_{pq}
  +\frac{1}{2}\sum_{pqrs}(2\rho_{pq}\rho_{rs} - \rho_{ps}\rho_{qr})I_{pqrs}G_{pqrs}
  + \frac{f^2_\alpha}{2}\sum_p (g_p - \eta_p)^2\rho_{pp}
  \nonumber \\ &
  +\frac{f^2_\alpha}{2} \sum_{pq}[2\rho_{pp}\rho_{qq}-\rho_{pq}\rho_{pq}](g_p - \eta_p)(g_q-\eta_q)
  + E_{res} + E_{ph}.
\end{align}
Here, $E_{res}$ accounts for the residual dipole self-energy:
\[
  E_{res} =\frac{(\Delta\lambda_\alpha)^2}{2}\bra{\Psi_e}(\boldsymbol{e}_\alpha\cdot\hat{\boldsymbol{D}})^2\ket{\Psi_e}.
\]
and $E_{ph}$ represents the photon energy:
\[
 E_{ph} = \sum_\alpha \omega_\alpha \cosh(2r)(n_\alpha+\frac{1}{2})
\]

The Fock matrix and gradients for the other variational parameters $\{\eta_p, f, r\}$ can be readily obtained by taking the functional derivative of $E$ with respect to the density matrix ($\rho_{pq}$) and the corresponding variational parameters.

\subsection{QEDHF with Variational Squeezed Gaussian State (SGS)}
\label{sec:qedhfSGS}

The unitary transformation for the Variational Squeezed Gaussian State (SGS) is:
\begin{equation}
    \hat{U} = \hat{S}(F)\hat{D}(\hat{f}).
\end{equation}
Hence, the transformed Hamiltonian is
\begin{equation}
\hat{\mathcal{H}}_{SGS} = \hat{D}^\dag(\hat{f}_\alpha)\left[\hat{S}^\dag(r)\hat{H}\hat{S}(r)\right]\hat{D}(\hat{f}_\alpha)
  \equiv \hat{D}^\dag(\hat{f}_\alpha)\hat{\mathcal{H}}_S\hat{D}(\hat{f}_\alpha),
\end{equation}
where the squeezed Hamiltonian, $\hat{\mathcal{H}}_S$, is given in Eq.~\ref{eq:hams}. The total Hamiltonian within the SGS transformation becomes,
\begin{align}
  \hat{\mathcal{H}}_{SGS} = &
  \hat{\mathcal{H}}_e(\hat{X})
  + \frac{1}{2}(\ldotd)^2
  + \sum_\alpha \sqrt{\frac{\omega_\alpha}{2}}(\ldotd)e^{-r}(\hb^\dag_\alpha + \hb_\alpha - 2\hat{f}_\alpha)
  \nonumber\\ &
  +
  \sum_\alpha\omega_\alpha \left\{\cosh(2r) (\hb^\dag_\alpha -\hat{f}_\alpha) (\hb_\alpha - \hat{f}_\alpha) + \sinh^2(r) -\sinh(2r)[(\hb^{\dag}_\alpha - \hat{f}_\alpha)^2 + (\hb_\alpha - \hat{f}_\alpha)^2] \right\}
  \nonumber\\ = &
  \hat{\mathcal{H}}_e(\hat{X})
  + \frac{1}{2}(\ldotd)^2
  - \sum_\alpha \sqrt{2\omega_\alpha}(\ldotd)e^{-r} \hat{f}_\alpha
  + \omega_\alpha\left[\cosh(2r) - 2\sinh(2r) \right]\hat{f}^2_\alpha
  \nonumber\\ &
  + \sum_\alpha \left\{\sqrt{\frac{\omega_\alpha}{2}}(\ldotd)e^{-r} -\omega_\alpha\left[\cosh(2r) -  2\sinh(2r)\right] \hat{f}_\alpha\right\}
  (\hb^\dag_\alpha + \hb_\alpha)
  \nonumber\\ &
  + \omega_\alpha\left\{\cosh(2r) \hb^\dag_\alpha\hb_\alpha + \sinh^2(r) -\sinh(2r)(\hb^{\dag 2}_\alpha + \hb^2_\alpha)
  \right\}.
\end{align}
This Hamiltonian is formally equivalent to Eq.~\ref{eq:hamcsc}, with $z_\alpha$ replaced by $\hat{f}_\alpha$ and $\hat{H}_e$ replaced by $\hat{\mathcal{H}}_e(\hat{X})$. After the transformation, the electronic Hamiltonian remains unaffected by the squeeze operator. Only the bilinear coupling and bosonic Hamiltonians are altered.
\textit{In the limit of a single bosonic Fock state (vacuum state), the bilinear term vanishes, and the optimization of the squeezed bosonic term becomes identical to the bare one. Consequently, the squeezing operator has no effect in the single Fock state limit.}

\subsection{QEDHF Formalism with Squeezed Coherent State (SCS)}
\label{sec:qedhfscs}

The squeezed coherent state (SCS) is defined as
\begin{equation}
    \ket{F, z} = \hat{U}\ket{0_p}= \hat{S}(F)\hat{D}(z)\ket{0_p}.
\end{equation}
The effective Hamiltonian in the squeezed coherent state (SCS) representation becomes,
\begin{align}
    \hat{H}^e_{scs} & = \hat{D}^\dag(z)\left[\hat{S}^\dag(F)\hat{H}_{PF}\hat{S}(F)\right]\hat{D}(z) 
    \equiv \hat{D}^\dag(z)\hat{\mathcal{H}}_{S}\hat{D}(z).
\end{align}
Since both $\hat{D}(z)$ and $\hat{S}(F)$ act only on the bosonic operator, the SCS transformation affects only the photonic part of the Hamiltonian.
By substituting $(\hb^\dag_\alpha - z^*_\alpha)$ and $(\hb_\alpha - z_\alpha)$ for $\hb^\dag_\alpha$ and $\hb_\alpha$, respectively, in Eq.~\ref{eq:hamcs}, the Hamiltonian in the SCS representation is given by:
\begin{align}
    \hat{\mathcal{H}}_{scs} = & \hat{H}_e + \frac{1}{2}(\ldotd)^2
    +\sqrt{\frac{\omega_\alpha}{2}}(\ldotd)
    e^{-r}(\hb^\dag_\alpha+\hb_\alpha - z_\alpha - z^*_\alpha)  + \omega_\alpha \cosh(2r)(\hb^\dag_\alpha - z^*_\alpha)\hb_\alpha - z_\alpha)
    \nonumber\\
    &~~~~~~~~~~~~~~~~~~~~~~
    -\omega_\alpha\sinh(2r)\left[(\hat{b}^{\dag}_\alpha - z^*_\alpha)^2 + (\hat{b}_\alpha - z_\alpha)^2\right] + \sinh^2(r) \omega_\alpha
    \nonumber\\
    = &\hat{H}_e + \frac{1}{2}(\ldotd)^2
    -\sqrt{2\omega_\alpha}{e^{-r}}(\ldotd)z_\alpha
    + \omega_\alpha \left[\cosh(2r) - 2\sinh(2r)\right]z^2_\alpha
    \nonumber\\
    & +\left\{\sqrt{\frac{\omega_\alpha}{2}}(\ldotd) e^{-r}
    - \omega_\alpha z_\alpha \left[\cosh(2r) - 2\sinh(2r)\right]\right\}(\hb^\dag_\alpha+\hb_\alpha)
    \nonumber\\
    & + \omega_\alpha \left\{\cosh(2r) \hat{b}^\dag_\alpha \hat{b}_\alpha + \sinh^2(r)
    -\sinh(2r)(\hat{b}^{\dagger 2}_\alpha+\hat{b}^2_\alpha)\right\}. \label{eq:hamcsc}
\end{align}

The new bilinear term becomes
\begin{equation}
    \hat{H}^{bi} = \left\{\sqrt{\frac{\omega_\alpha}{2}}(\ldotd) e^{-r}
    - \omega_\alpha z_\alpha \left[\cosh(2r) - 2\sinh(2r)\right]\right\}(\hb^\dag_\alpha + \hb_\alpha).
\end{equation}
The normalized photon Hamiltonian is:
\begin{equation}\label{eq:photonham}
 \hat{H}_{ph}=\omega_\alpha \cosh(2r) (\hat{b}^\dag_\alpha \hat{b}_\alpha + 1/2)
    -\omega_\alpha\sinh(2r)(\hat{b}^{\dagger 2}_\alpha+\hat{b}^2_\alpha).
\end{equation}
The DSE terms become
\[
   \hat{H}^{DSE} =\frac{1}{2}(\ldotd)^2
    -\sqrt{2\omega_\alpha}{e^{-r}}(\ldotd)z_\alpha
    + \omega_\alpha\left[\cosh(2r) - 2\sinh(2r)\right]  z^2_\alpha.
\]

Compared to the coherent state representation, the SCS transformation modifies only the photon Hamiltonian, as shown in Eq.~\ref{eq:photonham}, and boosts the DSE-mediated one-body integrals as
\begin{equation}
 h^{DSE}_{pq} = -\frac{1}{2}Q^\alpha_{pq} - e^{-r} \sqrt{\frac{\omega_\alpha}{2}}g_\alpha + \omega_\alpha \frac{z^2_\alpha}{N_e}\left[\cosh(2r) - 2\sinh(2r)\right] S_{pq},
\end{equation}
which is formally similar to the original DSE-mediated one electron integral in the CS representation in Eq.~\ref{eq:hamcs}.

\subsection{Coherent Squeeze State (CSS)}
\label{sec:qedhfCSS}

The Coherent Squeeze State (CSS) is defined as
\begin{equation}
    \hat{U}(z,F) = \hat{D}(z)\hat{S}(F)\ket{0}.
\end{equation}
Since $\hat{D}(z)$ and $\hat{S}(F)$ do not commute, the results from the Coherent Squeeze State and Squeezed Coherent State differ.

The effective Hamiltonian in the CSS representation can be obtained from Eq.~\ref{eq:hamcs} by replacing $\hb_\alpha$ with $\cosh(r)\hb_\alpha - \sinh(r)\hb^\dag_\alpha$,
\begin{align}
  \hat{H}_{CSS}
  = &
  \hat{H}_e + \frac{1}{2}(\ldotd)^2 +\sqrt{\frac{\omega_\alpha}{2}}e^{-r}[\boldsymbol{\lambda}_\alpha\cdot(\boldsymbol{D}-\langle\boldsymbol{D}\rangle)](\hb^\dag_\alpha+\hb_\alpha)
  + \frac{1}{2}\left[\boldsymbol{\lambda}_\alpha\cdot(\boldsymbol{D}-\langle\boldsymbol{D}\rangle)\right]^2
  \nonumber\\ &
  + \omega_\alpha \left\{\cosh(2r)(\hb^\dag_\alpha \hb_\alpha + \frac{1}{2}) -\sinh(2r)(\hb^{\dag 2}_\alpha + \hb^2_\alpha) \right\}.
\end{align}

The CSS transformation renormalizes only the bilinear coupling term and the photon Hamiltonian.
\textit{In the limit of a single-photon basis set (vacuum state), the bilinear term vanishes, and it is straightforward to verify that the squeezed state does not affect the ground state energy.}
However, in the many-photon basis set or the complete photon basis set limit, the bilinear term no longer vanishes, and the squeeze operation will impact the energies.

\section{Computation of Light-Matter Entanglement}
\label{smsec:entropy}

The wavefunction in the original basis set is
\begin{equation}
    \ket{\Phi} = \hat{U}\ket{\text{HF}}\otimes \ket{0_p}.
\end{equation}
From the mean-field solution to $\hat{\mathcal{H}}_{GSS}$, we obtain the $\ket{\text{HF}}$ wavefunction and the parameters $f_\alpha$ and $r_\alpha$. Hence, the original wavefunction can be constructed as
\begin{align}
  \ket{\Phi} = &
  e^{-\frac{f_\alpha\ldotd}{\sqrt{2\omega_\alpha}}(\hb -\hb^\dag)}\ket{\text{HF}} \otimes e^{-\frac{F}{2}(\hb^2-\hb^{\dag 2})}\ket{0_p}
  \nonumber\\
  = & e^{-\frac{f_\alpha\ldotd}{\sqrt{2\omega_\alpha}}(\hb -\hb^\dag)}\sum_\mu C_\mu\ket{\mu} \otimes e^{-\frac{F}{2}(\hb^2-\hb^{\dag 2})}\ket{0_p}
  \nonumber\\
  = & \sum_\mu C_\mu\ket{\mu} e^{-\left(\frac{f_\alpha\ldotd}{\sqrt{2\omega_\alpha}}\right)_\mu (\hb -\hb^\dag)}e^{-\frac{F}{2}(\hb^2-\hb^{\dag 2})}\ket{0_p}
  \nonumber\\
  \equiv &   \sum_\mu C_\mu\ket{\mu} \ket{z_\mu, F}.
\end{align}
where $\ket{z_\mu, F} \equiv e^{-\left(\frac{f_\alpha\ldotd}{\sqrt{2\omega_\alpha}}\right)_\mu (\hb -\hb^\dag)}e^{-\frac{F}{2}(\hb^2-\hb^{\dag 2})}\ket{0_p}$ is the squeezed coherent state.

The density operator is
\begin{equation}
    \rho = \ket{\Phi}\bra{\Phi} = \hat{U}\ket{\text{HF}, 0_p}\bra{0_p,\text{HF}}\hat{U}.
\end{equation}
Hence, in the original AO-Fock representation, the elements of the density matrix are
\begin{align}
    \rho^{mn}_{\mu\nu} = & \bra{\mu}\hat{U}\ket{\text{HF}, 0_p}\bra{0_p,\text{HF}}\hat{U}\ket{\nu} = \sum_i \bra{\mu}\hat{U}\ket{i, 0_p}\bra{0_p, i}\hat{U}\ket{\nu}
    \nonumber\\
    = &\sum_{i,\mu'\nu'} C^*_{\mu' i}C_{\nu'i}
    (A^\dag e^{-\frac{f_\alpha\boldsymbol{\eta}}{\sqrt{2\omega_\alpha}}} A)_{\mu\mu'}
    \hat{S}^\dag(F)\ket{\mu',0}\bra{\nu',0}\hat{S}^\dag(F)
    (A e^{-\frac{f_\alpha\boldsymbol{\eta}}{\sqrt{2\omega_\alpha}}} A^\dag)_{\nu'\nu}.
\end{align}

The photon density matrix can then be computed via the partial trace over the electronic DOF
\begin{equation}
  \rho_{ph} = \text{Tr}_e[\rho] = \sum_\mu |C_\mu|^2 \ket{Z_\mu, F}\bra{Z_\mu, F}.
\end{equation}
The corresponding Von Neumann entropy is
\begin{equation}
    S_{ph} = - K_B \text{Tr}\left[\rho_{ph}\ln(\rho_{ph})\right],
\end{equation}
which measures the light-matter entanglement.

\end{widetext}

\end{document}